\documentclass[12pt]{article}

\usepackage[margin=1in]{geometry}
\usepackage{graphicx}
\usepackage{booktabs}
\usepackage{amsmath,amssymb,amsthm}
\usepackage{natbib}
\usepackage{har2nat}
\usepackage{xurl}
\usepackage[colorlinks,linkcolor=red,citecolor=blue,breaklinks]{hyperref}
\hypersetup{bookmarksopen=true}
\usepackage[nameinlink]{cleveref}
\usepackage{caption}
\usepackage{subcaption}
\usepackage{float}
\usepackage{setspace}
\usepackage{enumitem}
\usepackage{pgfplots}
\pgfplotsset{compat=1.18}
\setlist{nosep}

\usepackage{textcomp}

\DeclareRobustCommand{\rnd}{\textcircled{r}}

\newcommand{\ATT}{\mathrm{ATT}}
\newcommand{\ATTst}{\ATT_{s,t}}
\newcommand{\ATTs}{\ATT_s}
\newcommand{\ATTgt}{\ATT_{g,t}}
\newcommand{\ATTg}{\ATT_g}
\newcommand{\ATTpt}{\ATT_{p,t}}
\newcommand{\ATTp}{\ATT_p}
\newtheorem{proposition}{Proposition}

\title{Which Policy Works, and Where? Estimation and Inference for State-Level Treatment Effects in Difference-in-Differences \thanks{We are grateful to the Canadian Institutes of Health Research (CIHR) for funding this project: grant number PJT-175079. We thank seminar participants at the 2025 CEA for helpful comments, and Eric Jamieson and Yunhan Liu for valuable research assistance.  

The authors declare that they have no relevant financial or non-financial competing interests to report. \par
\rnd Certified random order: A0Crar8Czf4g. }}
\author{Nichole Austin \rnd\, Sunny R. Karim \rnd\, Erin Strumpf \thanks{Austin: Dalhousie University, \url{nichole.austin@dal.ca}. Karim: Carleton University, \url{sunny.karim@cmail.carleton.ca}. Strumpf: McGill University, \url{erin.strumpf@mcgill.ca} (corresponding author).  Webb: Carleton University, \url{matt.webb@carleton.ca}.  } \rnd\, Matthew D. Webb}

\begin{document}
\maketitle

\begin{abstract}
Policies with a common objective and implementation date may differ in details or context. We distinguish the aggregate average treatment effect on the treated (ATT) from sub-aggregate ATTs defined by implementation cohort, jurisdiction, period, or policy type. UN-DID and DID-INT, two DiD estimators that construct jurisdiction-by-time effects, estimate these ATTs under parallel-trends conditions matched to the aggregation. In CPS placebo-law simulations, randomization inference is generally well-sized, though some jurisdiction-specific tests are conservative. The jackknife can be undefined for sub-aggregate ATTs; when defined, it over-rejects with few treated or comparison jurisdictions. Estimands and inference methods should match the policy question and implementation setting.
\end{abstract}

\onehalfspacing

\section{Introduction}

A large literature estimates policy impacts using variation in the timing of implementation across jurisdictions. Recent work has documented the weighting problems in two-way fixed effects estimation and developed alternative estimators that restrict comparisons to untreated or not-yet-treated jurisdictions \citep{goodman2021difference,de2020twott,callaway2021difference,sun2021estimating,borusyak2024revisiting}. In the Callaway--Sant'Anna framework, these comparisons identify cohort-by-time effects, where a cohort is the group of jurisdictions that implement a policy at the same time; these effects can then be aggregated across cohorts and periods \citep{callaway2021difference}. Policymakers often want to know the influence of specific policy design elements (e.g., a subsidized price vs. a reimbursement) and which contextual factors are facilitators or barriers to policy success. However, eligibility rules, benefit generosity, administration, enforcement, and context can differ across jurisdictions even when implementation timing is identical. Because cohort-by-time effects average across jurisdictions that implement at the same time, they combine these differences in policy content and context. The resulting cohort average may be well identified but may not answer the policy question of interest.

This paper contributes to two strands of the literature. One asks how to define and estimate interpretable treatment effects under staggered adoption \citep{goodman2021difference,callaway2021difference,sun2021estimating,borusyak2024revisiting}. The other asks how to do inference when treatment is assigned at the cluster level and the number of treated clusters is modest \citep{bertrand2004much,conley_2011,MACKINNON2020435,MNW-bootknife,karim2026improvedinferencecsdidusing}. Restricting comparisons to untreated or not-yet-treated jurisdictions does not by itself resolve the policy-content problem. When timing cohorts combine materially different interventions or policy environments, researchers should report jurisdiction-specific ATTs alongside cohort-specific and aggregate ATTs. Our contribution is twofold: we define and estimate aggregate and sub-aggregate ATTs that distinguish implementation cohort, jurisdiction, evaluation period, and policy type, and we study inference for these ATTs under staggered adoption.

The aggregate ATT averages across treated jurisdictions and post-treatment periods; sub-aggregate ATTs condition on implementation cohort, jurisdiction, evaluation period, policy type, or combinations of these dimensions. For example, states that raise the minimum wage in the same year may enact different percentage increases. A cohort ATT averages across those changes, whereas state-specific ATTs preserve the link between each statutory increase and its estimated effect. The policy-specific averages can combine jurisdictions that share a policy type even when they implement it in different periods. Our identification strategies retain the modern literature's explicit comparisons with never-treated or not-yet-treated jurisdictions, while the estimands align the aggregation with the implementation cohort, jurisdiction, evaluation period, or policy type of interest. The parallel-trends condition must align with the ATT being identified: jurisdiction-specific ATTs require jurisdiction-level parallel trends, whereas aggregate, cohort-specific, and policy-specific ATTs require the corresponding weighted parallel-trends conditions. We also maintain no anticipation and no spillovers across jurisdictions. Two recently developed DiD approaches, DiD with Unpoolable data (UN-DID) \citep{karim2024differenceindifferencesunpoolabledata} and Intersection DiD (DID-INT) \citep{karim2024good}, construct jurisdiction-by-time ATTs that can be aggregated to each of these levels. In the minimum-wage application, state-specific estimates vary substantially within implementation cohorts, including among states with similar statutory increases.

We also study inference for aggregate and sub-aggregate ATTs under staggered adoption. Using the recently developed UN-DID and DID-INT procedures \citep{karim2024differenceindifferencesunpoolabledata,karim2024good}, we compare randomization inference with the cluster jackknife \citep{MNW-bootknife,MNW-summclust, hansen2025jackknife, hansen2025standard}. In Current Population Survey (CPS)-based placebo-law simulations, RI rejection frequencies remain close to 5 percent for aggregate, cohort-specific, and policy-specific ATTs. For state-specific ATTs, RI is conservative when only one jurisdiction implements the placebo policy at each treatment date. When it remains defined, the jackknife follows a two-boundary pattern: it over-rejects when only a few treated jurisdictions contribute to the ATT and again when only a few comparison jurisdictions remain. The jackknife is undefined for cohort-specific ATTs when a cohort contains a single treated jurisdiction and for policy-specific ATTs when a policy type is represented by a single treated jurisdiction. The usual leave-one-jurisdiction-out jackknife is also undefined for jurisdiction-specific ATTs because deleting the focal jurisdiction removes it from the sample.

The number of treated and comparison jurisdictions is not the only consideration for inference with multi-stage DiD estimators. Deleting a jurisdiction can change aggregation weights, make the corresponding leave-one-jurisdiction-out estimate undefined, or require re-estimation of shared first-stage nuisance parameters. Whether deletion requires full re-estimation depends on how the first stage is specified. When DID-INT's covariate coefficients are jurisdiction specific, deleting another jurisdiction leaves them unchanged, so a Fast jackknife (FastJack) \citep{MNW-bootknife} calculation that holds first-stage estimates fixed can reproduce the leave-one-out estimator. When covariate coefficients or other nuisance parameters are shared across jurisdictions, full re-estimation is needed to reproduce the estimator in each leave-one-out sample.

\Cref{sec:policyquestion} uses Canadian in vitro fertilization  (IVF) and graduate-retention policies to show why timing-cohort averages may not answer the policy question. \Cref{sec:stateeffects} defines jurisdiction- and policy-specific ATTs and states their identifying assumptions. \Cref{sec:estimation} describes UN-DID and DID-INT and explains how aggregation and first-stage estimation affect jackknife inference. \Cref{sec:minwage} applies this framework to state minimum-wage changes. \Cref{sec:mc} presents the placebo-law simulation design and results. \Cref{sec:concl} concludes.
\section{Why Cohort Averages Can Miss the Policy Question}
\label{sec:policyquestion}

\begin{figure}[htbp]
\centering
\begin{subfigure}[t]{0.48\textwidth}
\centering
\includegraphics[width=\linewidth]{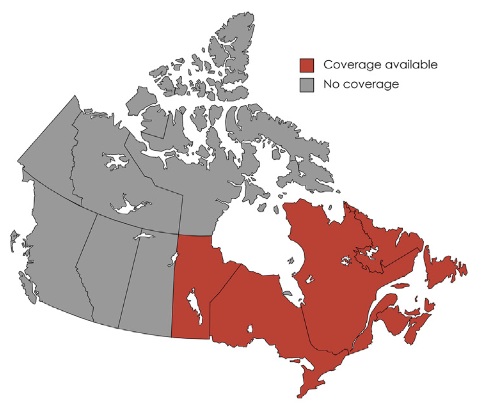}
\caption{Who is covered?}
\label{fig:who-treated}
\end{subfigure}
\hfill
\begin{subfigure}[t]{0.48\textwidth}
\centering
\includegraphics[width=\linewidth]{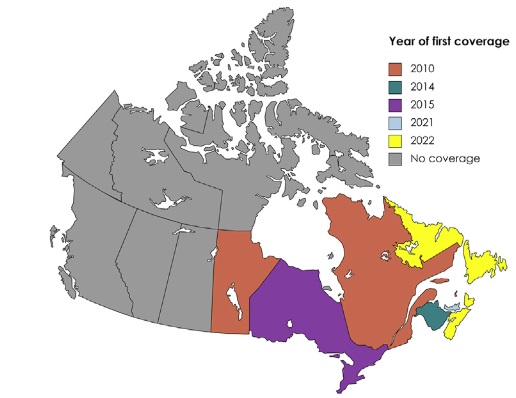}
\caption{When does coverage begin?}
\label{fig:when-treated}
\end{subfigure}

\vspace{0.5cm}

\begin{subfigure}[t]{0.48\textwidth}
\centering
\includegraphics[width=\linewidth]{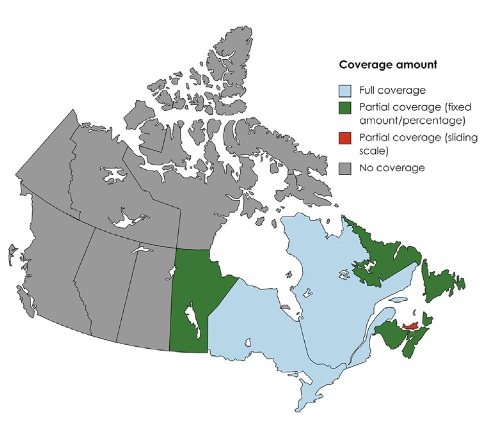}
\caption{How much support is offered?}
\label{fig:coverage-amount}
\end{subfigure}
\hfill
\begin{subfigure}[t]{0.48\textwidth}
\centering
\includegraphics[width=\linewidth]{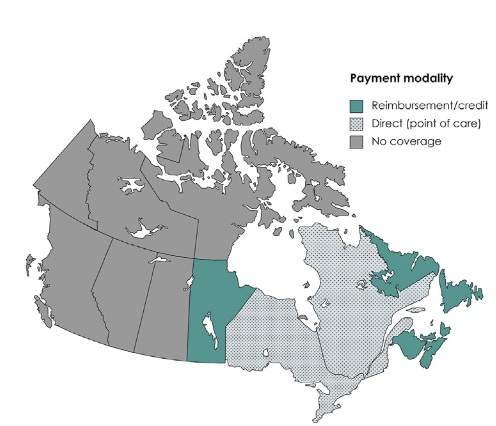}
\caption{How are benefits paid?}
\label{fig:benefit-payment}
\end{subfigure}
\caption{IVF coverage policies in Canada vary in eligibility, generosity, and payment design, so jurisdictions that begin coverage at the same time may implement different versions of the policy. Source: \citet{AustinApold2023}.}
\label{fig:ivf-coverage}
\end{figure}

Figure~\ref{fig:ivf-coverage} and Table~\ref{tab:grp.sum} illustrate the point with Canadian policy examples. The IVF subsidy programs studied in \citet{AustinApold2023} differ across Canadian provinces in who is covered, how much is paid, and how payments are delivered. The graduate-retention credits in \citet{mikola2023finish} differ in duration, refundability, and effective generosity. In both settings, implementation timing does not pin down treatment content. A timing cohort is therefore an econometric object, not necessarily a policy object of substantive interest.

\begin{table}[htbp]
\centering
\begin{tabular}{lcccc}
\toprule
\textbf{} & \multicolumn{4}{c}{\textbf{Program in Province}} \\
\textbf{} & \textbf{SK} & \textbf{MB} & \textbf{NS} & \textbf{NB} \\
\midrule
\textbf{start year} & \multicolumn{1}{c}{2007} & \multicolumn{1}{c}{2007} & \multicolumn{1}{c}{2006} & \multicolumn{1}{c}{2005} \\
\textbf{maximum amount} & \multicolumn{1}{c}{20k} & \multicolumn{1}{c}{25k} & \multicolumn{1}{c}{15k} & \multicolumn{1}{c}{20k} \\
\textbf{rebate per year} & \multicolumn{1}{c}{10\%, 20\%} & \multicolumn{1}{c}{4k, 10\%} & \multicolumn{1}{c}{2.5k} & \multicolumn{1}{c}{4k} \\
\textbf{NPV @ 5\%} & \multicolumn{1}{c}{16.9} & \multicolumn{1}{c}{14.1} & \multicolumn{1}{c}{13.3} & \multicolumn{1}{c}{12.6} \\
\textbf{refundable credit} & \multicolumn{1}{c}{Y*} & \multicolumn{1}{c}{N} & \multicolumn{1}{c}{N} & \multicolumn{1}{c}{N} \\
\textbf{rollover credit} & \multicolumn{1}{c}{N*} & \multicolumn{1}{c}{Y} & \multicolumn{1}{c}{N} & \multicolumn{1}{c}{Y} \\
\textbf{eligibility duration} & \multicolumn{1}{c}{7} & \multicolumn{1}{c}{10} & \multicolumn{1}{c}{6} & \multicolumn{1}{c}{20} \\
\textbf{application required} & \multicolumn{1}{c}{N} & \multicolumn{1}{c}{N} & \multicolumn{1}{c}{Y} & \multicolumn{1}{c}{Y} \\
\textbf{tuition based} & \multicolumn{1}{c}{Y} & \multicolumn{1}{c}{Y} & \multicolumn{1}{c}{N} & \multicolumn{1}{c}{Y} \\
\textbf{tuition \% refunded} & \multicolumn{1}{c}{100\%} & \multicolumn{1}{c}{60\%} & \multicolumn{1}{c}{---} & \multicolumn{1}{c}{50\%} \\
\textbf{program costs} & \multicolumn{1}{c}{35m} & \multicolumn{1}{c}{34m} & \multicolumn{1}{c}{25m} & \multicolumn{1}{c}{---} \\
\bottomrule
\end{tabular}
\caption{Graduate-retention programs implemented in the same year may nevertheless differ across provinces in duration, refundability, and generosity. Programs analyzed in \citet{mikola2023finish}.}
\label{tab:grp.sum}
\begin{minipage}{0.95\textwidth}
\footnotesize\textit{Notes:} An em dash denotes not applicable or not reported. The asterisks indicate Saskatchewan's announced 2012 change from a refundable credit to a rollover credit. If a refundable credit exceeds taxes owed, the excess is paid to the taxpayer rather than lost. An unused rollover credit can be carried forward and used in a later year.
\end{minipage}
\end{table}

Implementation timing can remain the main source of identifying variation even when policy content and context vary within a cohort. In such settings, cohort-specific ATTs should be interpreted alongside jurisdiction- or policy-specific ATTs that preserve the dimensions of policy variation relevant to the question.

\section{ATT Estimands and Identification}
\label{sec:stateeffects}

\subsection{Aggregate and Sub-Aggregate ATTs}
Let units $i=1,\dots,N$ be observed over periods $t=1,\dots,T$, and let jurisdictions be indexed by $s=1,\dots,H$. Let $S_i$ denote unit $i$'s jurisdiction, which is fixed over the study period, and let $D_{s,t}\in\{0,1\}$ denote jurisdiction-level treatment status, so $D_{i,t}=D_{S_i,t}$. Observed outcomes satisfy
\[
Y_{i,t}=D_{i,t}Y_{i,t}(1)+(1-D_{i,t})Y_{i,t}(0).
\]
Treatment is absorbing. If $G_s$ is the first treated period for jurisdiction $s$, with $G_s=\infty$ for never-treated jurisdictions, then jurisdiction $s$ is untreated before period $G_s$ and treated from period $G_s$ onward.

We use jurisdiction as the general term for the geographic unit at which a policy is implemented. In the U.S. application and CPS simulations, these jurisdictions are states, and we use state when referring specifically to those settings.

In the textbook two-jurisdiction, two-period DiD design, two-way fixed effects recovers the average treatment effect on the treated (ATT) under parallel trends. Under staggered adoption, $\beta^{\mathrm{TWFE}}$ from the same regression,
\begin{equation}
\label{eq:twfe}
Y_{i,t}=\alpha_i+\gamma_t+\beta^{\mathrm{TWFE}}D_{i,t}+u_{i,t},
\end{equation}
includes three comparisons: treated versus never treated, treated versus not-yet-treated, and treated versus already treated jurisdictions. With heterogeneous effects, some of these comparisons receive weights that are often hard to interpret and can even be negative \citep{goodman2021difference,de2020twott}. Researchers therefore do not want to include comparisons between treated and already-treated jurisdictions when estimating the ATT. These comparisons are commonly called ``forbidden comparisons'' \citep{goodman2021difference}. Some modern alternatives, notably \citet{callaway2021difference} solve that problem by estimating cohort-by-time ATTs and making the control group explicit.
If $\mathcal{G}_g=\{s:G_s=g\}$ is the set of jurisdictions first treated in period $g$, then a common building block is
\begin{equation}
\label{eq:attgt_intro}
\ATTgt=\mathbb{E}\!\left[Y_{i,t}(1)-Y_{i,t}(0)\mid S_i\in\mathcal{G}_g\right], \qquad t\ge g.
\end{equation}
Aggregate summaries then take the form
\begin{equation}
\label{eq:attg_att_intro}
\ATTg=\sum_{t=g}^{T}\omega_{g,t}\ATTgt,
\qquad
\ATT=\sum_{g:\mathcal{G}_g\ne\varnothing}\sum_{t=g}^{T}\Omega_{g,t}\ATTgt.
\end{equation}
Here $\omega_{g,t}\ge 0$ are post-treatment aggregation weights within cohort $g$, with $\sum_{t=g}^{T}\omega_{g,t}=1$, and $\Omega_{g,t}\ge 0$ are the overall aggregation weights with \(\sum_{g:\mathcal{G}_g\ne\varnothing}\sum_{t=g}^{T}\Omega_{g,t}=1\).

A calendar-time ATT can similarly be formed by averaging across treated cohorts observed in period \(t\),

$$ \ATT_t = \sum_{g\le t}\rho_{g\mid t}\ATTgt, \qquad \rho_{g\mid t}\ge 0, \qquad \sum_{g\le t}\rho_{g\mid t}=1, $$
where \(\rho_{g\mid t}\) are the aggregation weights across treated cohorts observed in period \(t\). Thus \(\ATTg\) holds the implementation cohort fixed and averages over post-treatment periods, whereas \(\ATT_t\) holds the evaluation period fixed and averages over treated cohorts.

Unlike TWFE, these estimators exclude already-treated jurisdictions from the comparison group, eliminating the negative weights generated by those comparisons. They also make the comparison group and aggregation weights explicit. However, they still organize effects by implementation date rather than by policy content. If two jurisdictions implement policies in the same year but one offers a generous refundable credit and the other offers a small nonrefundable subsidy with short eligibility, the average effect for that timing cohort may not match any policy a reader cares about. The problem is not averaging by itself; it is averaging across jurisdictions whose policies differ in design or implementation.

For a treated jurisdiction $s$, let $\mathcal T_s=\{t:G_s\le t\le T\}$ denote its observed post-treatment periods. The natural jurisdiction-specific building block is
\begin{equation}
\label{eq:attst_def}
\ATTst
:=
\mathbb{E}\!\left[Y_{i,t}(1)-Y_{i,t}(0)\mid S_i=s\right],
\qquad t\in\mathcal T_s.
\end{equation}
This object answers the question, ``What was the effect in jurisdiction $s$ at time $t$?'' Averaging over $\mathcal{T}_s$, with weights $\nu_{s,t}\ge 0$ and $\sum_{t\in\mathcal{T}_s}\nu_{s,t}=1$, gives the jurisdiction-specific average ATT
\begin{equation}
\label{eq:atts_def}
\ATTs=\sum_{t\in\mathcal{T}_s}\nu_{s,t}\ATTst.
\end{equation}
The weights define the temporal aggregation: equal weights average across periods, while sample-size weights average across treated observations.

The relation to cohort effects is straightforward. If $\mathcal{G}_g$ is the timing cohort that first implements the policy in period $g$, then
\begin{equation}
\label{eq:attgt_as_avg_attst}
\ATTgt
=
\sum_{s\in\mathcal{G}_g} \pi_{s\mid g,t}\ATTst,
\qquad t\ge g.
\end{equation}
Here $\pi_{s\mid g,t}$ is the share of treated units in jurisdiction $s$ among treated units in cohort $g$ at time $t$, so $\sum_{s\in\mathcal{G}_g}\pi_{s\mid g,t}=1$; these weights are often proportional to the number of observations in each jurisdiction-by-time cell. Many estimators report $\ATTgt$ without retaining the jurisdiction-specific estimates, although \cref{eq:attgt_as_avg_attst} shows that the cohort-time ATT averages $\ATTst$ over jurisdictions first treated in the same period. This cohort-specific ATT is useful when those jurisdictions implement comparable policies, but it masks within-cohort variation.

Between cohort-specific and jurisdiction-specific ATTs are policy-specific ATTs. Let $P_s$ denote the policy type implemented by jurisdiction $s$, and let $p$ index its possible values. For example, $P_s=A$ might denote a direct subsidy and $P_s=B$ a refundable tax credit, while $P_s=0$ for never-treated jurisdictions.

For policy type $p$ and calendar period $t$, define the set of contributing treated jurisdictions
\[
\mathcal{S}_{p,t}=\{s:P_s=p,\ G_s\le t\}.
\]
When this set is nonempty, the policy-specific ATT at time $t$ is
\begin{equation}
\label{eq:attpt_def}
\ATTpt
=
\sum_{s\in\mathcal{S}_{p,t}} \pi_{s\mid p,t}\ATTst,
\qquad
\sum_{s\in\mathcal{S}_{p,t}}\pi_{s\mid p,t}=1.
\end{equation}
Aggregating over the post-treatment periods in which policy type $p$ is observed gives
\begin{equation}
\label{eq:attp_def}
\ATTp
=
\sum_{t\in\mathcal{T}_{p}}\nu_{p,t}\ATTpt,
\qquad
\sum_{t\in\mathcal{T}_{p}}\nu_{p,t}=1.
\end{equation}
All of these estimands are ATTs. The aggregate $\ATT$ averages across treated jurisdictions and post-treatment periods; sub-aggregate ATTs condition on cohort ($\ATTg$ and $\ATTgt$), jurisdiction ($\ATTs$ and $\ATTst$), or policy type ($\ATTp$ and $\ATTpt$). The time-indexed estimands also condition on the evaluation period.

\subsection{Identification}
The same distinction appears in the identifying assumptions. Let
\[
\bar{Y}^{0}_{s,t}=\mathbb{E}[Y_{i,t}(0)\mid S_i=s],
\qquad
\bar{Y}^{0}_{g,t}=\sum_{s\in\mathcal{G}_g}\pi_{s\mid g,t}\bar{Y}^{0}_{s,t},
\qquad
\bar{Y}^{0}_{p,t}=\sum_{s\in\mathcal{S}_{p,t}}\pi_{s\mid p,t}\bar{Y}^{0}_{s,t}.
\]
For treated jurisdiction $s$ and post-treatment period $t$, let $t'_s=G_s-1$ and let $\mathcal{C}_{s,t}$ denote the comparison set used for the jurisdiction-time contrast. This set consists of never-treated jurisdictions and, when allowed by the estimator, jurisdictions not yet treated by $t$. Define the comparison trend
\[
\Delta\bar{Y}^{C}_{s,t}
=
\sum_{s'\in\mathcal{C}_{s,t}}\lambda_{s'\mid s,t}
\left(\bar{Y}_{s',t}-\bar{Y}_{s',t'_s}\right),
\qquad
\sum_{s'\in\mathcal{C}_{s,t}}\lambda_{s'\mid s,t}=1.
\]
The same comparison jurisdictions are evaluated at $t$ and $t'_s$. Because every jurisdiction in $\mathcal C_{s,t}$ is untreated in both periods, $\Delta\bar Y^C_{s,t}$ is a weighted untreated comparison trend.

\begin{figure}[htbp]
\centering

\begin{subfigure}[t]{0.23\textwidth}
\centering
\begin{tikzpicture}
\begin{axis}[
title=State 1 (Treated),
xlabel=Time,
ylabel=Earnings,
width=0.95\linewidth,
height=4cm,
ymin=-1, ymax=12,
xmin=-10, xmax=0,
xtick={-10,-5,0},
ytick={-1,0,2,4,6,8,10,12},
grid=major
]
\addplot[thick, blue, mark=*] coordinates {
(-10,1) (-9,0) (-8,3) (-7,2) (-6,5) (-5,4) (-4,7) (-3,6) (-2,9) (-1,8) (0,11)
};
\end{axis}
\end{tikzpicture}
\end{subfigure}\hfill%
\begin{subfigure}[t]{0.23\textwidth}
\centering
\begin{tikzpicture}
\begin{axis}[
title=State 2 (Treated),
xlabel=Time,
ylabel=,
width=0.95\linewidth,
height=4cm,
ymin=-1, ymax=12,
xmin=-10, xmax=0,
xtick={-10,-5,0},
ytick=\empty,
grid=major
]
\addplot[thick, red, mark=*] coordinates {
(-10,-1) (-9,2) (-8,1) (-7,4) (-6,3) (-5,6) (-4,5) (-3,8) (-2,7) (-1,10) (0,9)
};
\end{axis}
\end{tikzpicture}
\end{subfigure}\hfill%
\begin{subfigure}[t]{0.23\textwidth}
\centering
\begin{tikzpicture}
\begin{axis}[
title=State 3 (Control),
xlabel=Time,
ylabel=,
width=0.95\linewidth,
height=4cm,
ymin=-1, ymax=12,
xmin=-10, xmax=0,
xtick={-10,-5,0},
ytick=\empty,
grid=major
]
\addplot[thick, green!60!black, mark=*] coordinates {
(-10,0) (-9,1.1) (-8,2.2) (-7,3.0) (-6,4.1) (-5,5.1) (-4,6.0) (-3,7.1) (-2,8.2) (-1,9.0) (0,10.1)
};
\end{axis}
\end{tikzpicture}
\end{subfigure}\hfill%
\begin{subfigure}[t]{0.23\textwidth}
\centering
\begin{tikzpicture}
\begin{axis}[
title=Treated Cohort,
xlabel=Time,
ylabel=,
width=0.95\linewidth,
height=4cm,
ymin=-1, ymax=12,
xmin=-10, xmax=0,
xtick={-10,-5,0},
ytick=\empty,
grid=major
]
\addplot[thick, black, dashed, mark=*] coordinates {
(-10,0) (-9,1) (-8,2) (-7,3) (-6,4) (-5,5) (-4,6) (-3,7) (-2,8) (-1,9) (0,10)
};
\end{axis}
\end{tikzpicture}
\end{subfigure}

\caption{A case in which offsetting pre-trends can make the treated cohort average look as if the parallel trends assumption holds even though individual treated states do not satisfy state-level parallel trends.}
\label{fig:offsetting-shocks}
\end{figure}
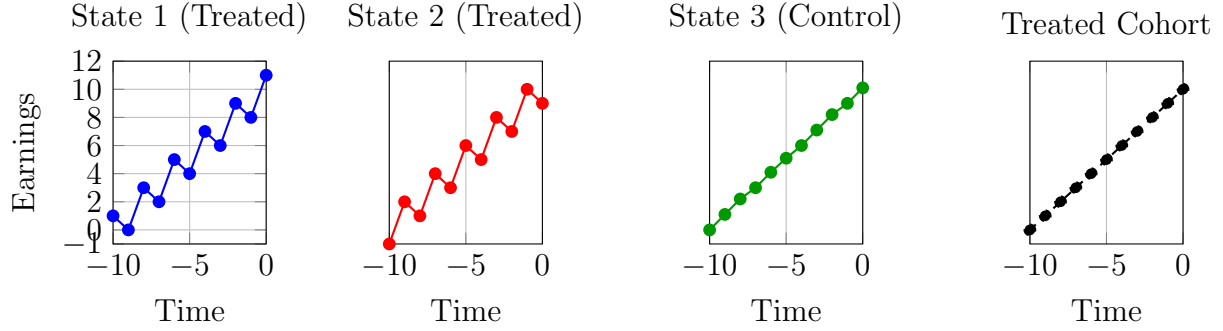

The weights $\lambda_{s'\mid s,t}$ are the comparison weights used by the estimator for the corresponding contrast. With no anticipation, and no cross-jurisdiction spillovers, identification of $\ATTgt$ requires parallel trends for the cohort average:
\begin{equation}
\label{eq:pt_cohort}
\sum_{s\in\mathcal{G}_g}\pi_{s\mid g,t}
\mathbb{E}\!\left[\bar{Y}^{0}_{s,t}-\bar{Y}^{0}_{s,t'_s}\right]
=
\sum_{s\in\mathcal{G}_g}\pi_{s\mid g,t}
\mathbb{E}\!\left[\Delta\bar{Y}^{C}_{s,t}\right],
\qquad t\ge g,
\end{equation}
This condition requires the weighted untreated trend for jurisdictions first treated in period $g$ to match the weighted comparison-jurisdiction trend. It can hold even when individual jurisdictions in $\mathcal G_g$ violate jurisdiction-specific parallel trends because their deviations may offset, as \cref{fig:offsetting-shocks} illustrates.
Identification of $\ATTpt$ can be based on an analogous policy-specific average condition. Because a policy type can appear in more than one timing cohort, the pre-treatment period $t'_s$ may differ across jurisdictions in $\mathcal S_{p,t}$. The direct policy-specific parallel-trends condition is
\begin{equation}
\label{eq:pt_policy}
\sum_{s\in\mathcal{S}_{p,t}}\pi_{s\mid p,t}
\mathbb{E}\!\left[\bar{Y}^{0}_{s,t}-\bar{Y}^{0}_{s,t'_s}\right]
=
\sum_{s\in\mathcal{S}_{p,t}}\pi_{s\mid p,t}
\mathbb{E}\!\left[\Delta\bar{Y}^{C}_{s,t}\right],
\qquad t\in\mathcal{T}_{p}.
\end{equation}
This condition requires the weighted untreated trend for jurisdictions implementing policy type $p$ to match the corresponding weighted comparison-jurisdiction trend. It is weaker than requiring every jurisdiction in $\mathcal{S}_{p,t}$ to satisfy the same restriction on its own and is the direct identifying condition for $\ATTpt$. It likewise permits offsetting jurisdiction-specific violations.

Identification of $\ATTst$ requires the stronger jurisdiction-level condition
\begin{equation}
\label{eq:pt_state}
\mathbb{E}\!\left[\bar{Y}^{0}_{s,t}-\bar{Y}^{0}_{s,t'_s}\right]
=
\mathbb{E}\!\left[\Delta\bar{Y}^{C}_{s,t}\right],
\qquad t\ge G_s,
\end{equation}
and $\ATTs$ inherits that requirement because it aggregates $\ATTst$. If the jurisdiction-level condition holds for each $s\in\mathcal{S}_{p,t}$, then $\ATTpt$ is also identified as a weighted average of identified jurisdiction-time effects, and $\ATTp$ is identified by aggregating those policy-time effects. For the policy-specific aggregate, however, \cref{eq:pt_policy} is the direct identifying assumption.

\begin{proposition}
\label{prop:att-identification}
Fix a treated jurisdiction $s$ and a post-treatment period $t\ge G_s$. Treatment is absorbing, so once jurisdiction $s$ is treated it remains treated thereafter. Suppose also that (i) no anticipation holds, so $Y_{i,t'_s}=Y_{i,t'_s}(0)$ among units in jurisdiction $s$; (ii) no cross-jurisdiction spillovers hold, so outcomes for jurisdictions in $\mathcal{C}_{s,t}$ are untreated potential outcomes; and (iii) the jurisdiction-level parallel-trends condition in \cref{eq:pt_state} holds. Then the DiD contrast
\[
\left(\bar{Y}_{s,t}-\bar{Y}_{s,t'_s}\right)
-
\Delta\bar{Y}^{C}_{s,t}
\]
identifies $\ATTst$. If \cref{eq:pt_state} holds for every $t\in\mathcal{T}_s$, then the weighted average of these identified contrasts identifies $\ATTs$.
\end{proposition}

By no anticipation, $\bar{Y}_{s,t'_s}=\bar{Y}^{0}_{s,t'_s}$. By no spillovers, $\Delta\bar{Y}^{C}_{s,t}$ is an untreated comparison trend. Subtracting the jurisdiction-level parallel-trends restriction from the observed jurisdiction trend leaves $\mathbb{E}[Y_{i,t}(1)-Y_{i,t}(0)\mid S_i=s]$. Aggregating over $t$ with weights $\nu_{s,t}$ gives $\ATTs$. Similarly, subtracting \cref{eq:pt_policy} from the observed policy-specific average trend identifies $\ATTpt$, and aggregating those identified policy-by-time effects gives $\ATTp$. Finally, if \cref{eq:pt_state} holds for every $s\in\mathcal{G}_g$, then multiplying by $\pi_{s\mid g,t}$ and summing over jurisdictions implies \cref{eq:pt_cohort}, so the cohort-time estimand $\ATTgt$ is identified as a weighted average of identified jurisdiction-time effects. The same implication holds for \cref{eq:pt_policy} when the jurisdiction-level condition holds for every jurisdiction in $\mathcal{S}_{p,t}$. The converse need not hold. One jurisdiction in the cohort or policy-type group can drift upward relative to comparisons while another drifts downward by a similar amount, so the weighted average deviation is close to zero even though neither treated jurisdiction satisfies jurisdiction-level parallel trends on its own.

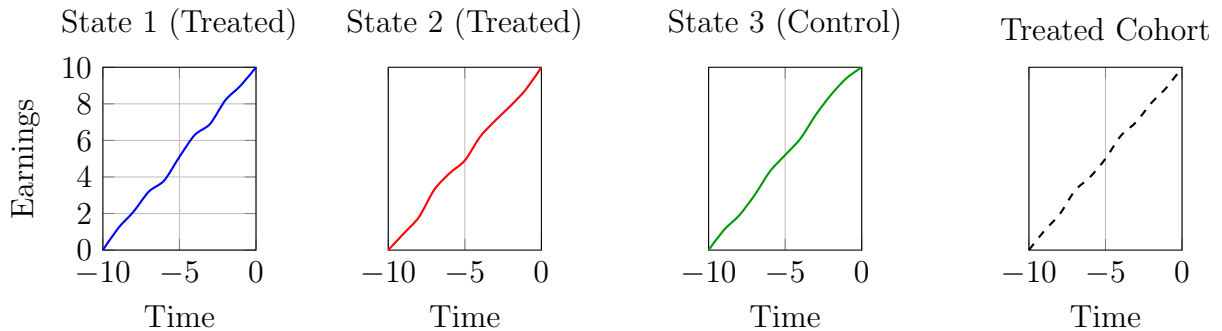
\begin{figure}[htbp]
\centering

\begin{subfigure}[t]{0.23\textwidth}
\centering
\begin{tikzpicture}
\begin{axis}[
title=State 1 (Treated),
xlabel=Time,
ylabel=Earnings,
width=0.95\linewidth,
height=4cm,
ymin=0, ymax=10,
xmin=-10, xmax=0,
xtick={-10,-5,0},
ytick={0,2,4,6,8,10},
grid=major
]
\addplot[smooth, thick, blue] coordinates {
(-10,0) (-9,1.2) (-8,2.1) (-7,3.2) (-6,3.8) (-5,5.1) (-4,6.3) (-3,6.9) (-2,8.2) (-1,9.0) (0,10)
};
\end{axis}
\end{tikzpicture}
\end{subfigure}\hfill%
\begin{subfigure}[t]{0.23\textwidth}
\centering
\begin{tikzpicture}
\begin{axis}[
title=State 2 (Treated),
xlabel=Time,
ylabel=,
width=0.95\linewidth,
height=4cm,
ymin=0, ymax=10,
xmin=-10, xmax=0,
xtick={-10,-5,0},
ytick=\empty,
grid=major
]
\addplot[smooth, thick, red] coordinates {
(-10,0) (-9,0.9) (-8,1.8) (-7,3.3) (-6,4.2) (-5,4.9) (-4,6.2) (-3,7.1) (-2,7.9) (-1,8.8) (0,10)
};
\end{axis}
\end{tikzpicture}
\end{subfigure}\hfill%
\begin{subfigure}[t]{0.23\textwidth}
\centering
\begin{tikzpicture}
\begin{axis}[
title=State 3 (Control),
xlabel=Time,
ylabel=,
width=0.95\linewidth,
height=4cm,
ymin=0, ymax=10,
xmin=-10, xmax=0,
xtick={-10,-5,0},
ytick=\empty,
grid=major
]
\addplot[smooth, thick, green!60!black] coordinates {
(-10,0) (-9,1.1) (-8,1.9) (-7,3.0) (-6,4.3) (-5,5.2) (-4,6.1) (-3,7.4) (-2,8.5) (-1,9.4) (0,10)
};
\end{axis}
\end{tikzpicture}
\end{subfigure}\hfill%
\begin{subfigure}[t]{0.23\textwidth}
\centering
\begin{tikzpicture}
\begin{axis}[
title=Treated Cohort,
xlabel=Time,
ylabel=,
width=0.95\linewidth,
height=4cm,
ymin=0, ymax=10,
xmin=-10, xmax=0,
xtick={-10,-5,0},
ytick=\empty,
grid=major
]
\addplot[smooth, thick, black, dashed] coordinates {
(-10,0) (-9,1.05) (-8,1.95) (-7,3.25) (-6,4.00) (-5,5.00) (-4,6.25) (-3,7.00) (-2,8.05) (-1,8.90) (0,10)
};
\end{axis}
\end{tikzpicture}
\end{subfigure}

\caption{A case in which the state-level and cohort-level parallel trends assumptions are both plausible. States 1 and 2 are treated at the same time and can be averaged into a treated cohort.}
\label{fig:parallel-ok}
\end{figure}

\Cref{fig:parallel-ok,fig:offsetting-shocks} illustrate the difference. In both figures, States 1 and 2 are treated at the same time and therefore belong to the same treated cohort, while State 3 is the untreated control. The fourth panel plots the treated-cohort trend, defined as the average of States 1 and 2. In \Cref{fig:parallel-ok}, States 1 and 2 each move approximately in parallel with State 3 before treatment, so both the state-level and cohort-level parallel-trends restrictions are plausible. In \Cref{fig:offsetting-shocks}, State 1 drifts upward relative to State 3 while State 2 drifts downward by a similar amount overall.  The slope reverses each period, representing a pattern of seasonality or cyclicality. Averaging States 1 and 2 therefore produces a treated-cohort trend that looks close to the control trend, even though neither treated state satisfies state-level parallel trends. That illustrates the paper's central trade-off. State-level estimands are often more interpretable, but they require stronger assumptions for identification.

These examples also have an implication for pre-trend diagnostics. Cohort-level pre-treatment trends can appear parallel even when jurisdiction-specific pre-treatment trends show offsetting departures. Graphical and other pre-trend diagnostics should therefore be examined at the same level of aggregation as the ATT of interest.

\section{Estimation and Inference}
\label{sec:estimation}

We do not propose a new estimator. Instead, we use two recently developed procedures, UN-DID and DID-INT, to estimate the aggregate and sub-aggregate ATTs defined above and study inference for them.  UN-DID is designed for settings in which microdata cannot be pooled: jurisdiction-by-time means and long differences are computed within data silos and then combined \citep{karim2024differenceindifferencesunpoolabledata}. DID-INT is designed for pooled data and permits flexible covariate adjustment, including specifications in which covariate coefficients vary across jurisdictions, over time, or both \citep{karim2024good}.  Both estimators rely on multi-stage estimation which is explained in Section \ref{sec:multi}.

Because treatment is assigned at the jurisdiction level and outcomes may be correlated within jurisdictions, inference should account for jurisdiction-level clustering \citep{bertrand2004much, MNW-guide}. Conventional jurisdiction-clustered inference is challenging for both estimators. In UN-DID, microdata are assumed to remain siloed at the jurisdiction level, so conventional cluster-robust standard errors at the jurisdiction level are unavailable in the first stage because there is only one cluster per silo. In DID-INT, the flexible first-stage specification can include jurisdiction-specific time effects and covariate coefficients. The regressors associated with these nuisance parameters are nonzero within only one jurisdiction. Additionally, the one-way cluster-robust covariance matrix has rank at most \(H-1\), making conventional clustered inference for this set of jurisdiction-specific parameters infeasible. We are concerned with inference on the aggregate and sub-aggregate ATTs rather than the first-stage nuisance parameters themselves. The issue is that uncertainty from first-stage estimation contributes to the sampling variation of the ATT and therefore must be accounted for in inference on the ATT. We consequently consider randomization inference and the cluster jackknife. Care must be taken as the jackknife is available only when the relevant ATT remains defined after each jurisdiction is deleted.  It must also be implemented in a way that correctly accounts for any first-stage nuisance parameters that change after removing a jurisdiction.

\subsection{The CCC assumption}
\label{subsec:ccc}

The key additional ingredient for DID-INT is the common causal covariates (CCC) assumption. Time-varying covariates may affect outcomes differently across jurisdictions or periods. If the first-stage adjustment incorrectly imposes common covariate effects, the residualized outcomes can violate the relevant parallel-trends condition even when a more flexible adjustment would satisfy it \citep{karim2024good}. DID-INT accommodates such differences by allowing the covariate function \(f(X_{i,s,t})\) to vary across jurisdictions, periods, or both.

Four DID-INT specifications are relevant. The homogeneous specification imposes common covariate effects across jurisdictions and periods and is appropriate when two-way CCC holds. The jurisdiction-varying and time-varying specifications allow covariate effects to differ across jurisdictions and periods, respectively. The two-way specification allows them to differ across both dimensions and accommodates violations of two-way CCC. UN-DID naturally allows covariate effects to vary across jurisdictions; adding time interactions extends it to the two-way case.

These distinctions matter for two reasons. First, they determine which identifying restriction on residualized untreated outcomes is plausible. Second, they determine whether deleting one jurisdiction changes nuisance parameters used for the remaining jurisdictions, which is why the CCC discussion is central for the jackknife below.

\subsection{A multi-stage estimator architecture}
\label{sec:multi}

Both UN-DID and DID-INT can be viewed as multi-stage procedures. First, the researcher estimates nuisance objects needed to form untreated counterfactual trends, such as outcome regressions, adjustment terms, weights, or covariate residualizations. In the implementations studied here, the basic ingredients are long differences of the form \(Y_{s,t}-Y_{s,t'}\) for treated jurisdictions and corresponding differences for the comparison jurisdictions, where \(t'=G_s-1\) is the period immediately before treated jurisdiction $s$ implements the policy. Second, these objects are used to construct treated-jurisdiction-by-time building blocks $\widehat{\ATT}_{s,t}$ for treated jurisdictions $s$ and post-treatment periods $t\ge G_s$. Third, the building blocks are aggregated to form the relevant aggregate or sub-aggregate ATT.

The jurisdiction-specific average is
\[
\widehat{\ATT}_s=\sum_{t\in\mathcal{T}_s}\nu_{s,t}\widehat{\ATT}_{s,t},
\]
where equal weights over the post-treatment periods are a common default. 
Cohort-level objects are recovered by averaging within timing cohorts,
\[
\widehat{\ATT}_{g,t}=\sum_{s\in\mathcal{G}_g}\pi_{s\mid g,t}\widehat{\ATT}_{s,t},
\qquad
\widehat{\ATT}_{g}=\sum_{t\in\mathcal{T}_g}\omega_{g,t}\widehat{\ATT}_{g,t},
\]
with \(\pi_{s\mid g,t}\) often proportional to the number of observations in each jurisdiction-time cell. In estimators such as Callaway--Sant'Anna \citep{callaway2021difference}, the jurisdiction label is not retained as part of the reported estimand once observations are grouped into a treatment-timing cohort. Policy-specific ATTs are recovered by averaging the same jurisdiction-time building blocks within policy type,
\[
\widehat{\ATT}_{p,t}=\sum_{s\in\mathcal{S}_{p,t}}\pi_{s\mid p,t}\widehat{\ATT}_{s,t},
\qquad
\widehat{\ATT}_{p}=\sum_{t\in\mathcal{T}_p}\nu_{p,t}\widehat{\ATT}_{p,t},
\]
while an overall summary takes the form
\[
\widehat{\ATT}
=
\sum_{s:G_s<\infty}\sum_{t\in\mathcal{T}_s}\Omega_{s,t}\widehat{\ATT}_{s,t},
\qquad
\Omega_{s,t}\ge 0,
\qquad
\sum_{s,t}\Omega_{s,t}=1.
\]

This architecture separates the object being estimated from the procedure used to estimate it. The same building blocks can produce jurisdiction-level, cohort-level, or aggregate summaries. But inference depends on all three stages. A resampling procedure may change the building blocks, the aggregation weights, or whether the corresponding leave-one-jurisdiction-out estimate is defined.

Different weights also answer different questions. Equal weights over treated jurisdiction-by-time cells emphasize treated exposure. Equal weights over jurisdictions emphasize the typical treated jurisdiction. Population or treated-count weights emphasize aggregate policy reach.  Under equal weighting of treated jurisdiction-by-time cells, early adopters also receive more total weight because they contribute more post-treatment periods. Researchers should therefore defend the aggregation rule as part of the design rather than treat it as a bookkeeping detail.

\subsection{When leave-one-jurisdiction-out estimates are defined}

The cluster jackknife deletes one jurisdiction at a time and estimates variance from the dispersion of the leave-one-jurisdiction-out replicates,
\begin{equation}
\label{eq:jk_var}
\widehat{\mathrm{Var}}_{\mathrm{JK}}(\widehat{\theta})
=
\frac{H-1}{H}\sum_{s=1}^{H}\left(\widehat{\theta}^{(s)}-\bar{\theta}^{(\cdot)}\right)^2,
\qquad
\bar{\theta}^{(\cdot)}=\frac{1}{H}\sum_{s=1}^{H}\widehat{\theta}^{(s)}.
\end{equation}
The method is appealing when treatment is assigned at the jurisdiction level and the ATT remains defined after each jurisdiction is deleted \citep{MNW-bootknife,karim2026improvedinferencecsdidusing}.

This condition is not automatic in staggered-adoption designs. An aggregate ATT can usually be recomputed after deleting one jurisdiction, although its weights must be renormalized. A cohort-specific ATT can be recomputed only if the relevant cohort remains nonempty. If a cohort contains one treated jurisdiction, deleting it makes the corresponding leave-one-out estimate of $\ATTg$ or $\ATTgt$ undefined. The same condition applies to policy-specific ATTs: deleting one jurisdiction must leave at least one treated jurisdiction contributing to the relevant $\ATTp$ or $\ATTpt$. If a policy type or policy-by-time cell is represented by one treated jurisdiction, the standard delete-one-jurisdiction jackknife cannot be computed for that ATT.

Jurisdiction-specific ATTs are different. After deleting jurisdiction $s$, the data needed to estimate $\ATTs$ and $\ATTst$ are absent from the leave-one-jurisdiction-out sample. The standard delete-one-jurisdiction jackknife therefore cannot be computed for these ATTs. Other restricted deletion schemes are conceivable, but they represent a different sampling experiment. The simulation results below therefore report randomization inference for jurisdiction-specific ATTs and use the jackknife only when every required leave-one-out estimate is defined.

\subsection{A four-jurisdiction example: jackknife reweighting under staggered adoption}

A four-jurisdiction example shows that deletion can change not only whether a leave-one-out estimate is defined but also the composition and weights of the ATT being estimated. Consider two never-treated comparisons $c_1$ and $c_2$, one early-treated jurisdiction $j$, and one late-treated jurisdiction $\ell$. There are three periods, $t=0,1,2$. The early-treated jurisdiction first implements the policy in period 1, so $G_j=1$; the late-treated jurisdiction first implements the policy in period 2, so $G_\ell=2$; and $G_{c_1}=G_{c_2}=\infty$. The comparison group consists only of $c_1$ and $c_2$; the late-treated jurisdiction $\ell$ is not used as a not-yet-treated comparison in period 1. The early-treated jurisdiction contributes two post-treatment cells, $(j,1)$ and $(j,2)$, while the late-treated jurisdiction contributes one post-treatment cell, $(\ell,2)$.

Suppose the reported aggregate gives equal weight to treated jurisdiction-time cells:
\[
\ATT^{\mathrm{avg}}
=
\frac{1}{3}\left(\ATT_{j,1}+\ATT_{j,2}+\ATT_{\ell,2}\right).
\]
Deleting a comparison jurisdiction leaves all three treated cells used to form the aggregate ATT, although the comparison aggregate used to estimate counterfactual trends changes. Deleting the late-treated jurisdiction instead leaves
\[
\ATT^{(\ell)}
=
\frac{1}{2}\left(\ATT_{j,1}+\ATT_{j,2}\right),
\]
while deleting the early-treated jurisdiction leaves
\[
\ATT^{(j)}
=
\ATT_{\ell,2}.
\]
The leave-one-jurisdiction-out estimates differ not only because of sampling variation. When effects vary across jurisdictions or policy types, deleting a jurisdiction changes the substantive mix of treatment-effect cells being averaged. Deleting the early-treated jurisdiction removes two treated cells, while deleting the late-treated jurisdiction removes one. Jackknife dispersion therefore reflects both effect heterogeneity and reweighting induced by staggered exposure lengths.

This example does not establish over-rejection by itself, but it highlights why jackknife inference can be unreliable when only a few treated jurisdictions contribute to a cohort-specific ATT \citep{conley_2011,mackinnon2017wild,MACKINNON2020435}. Deleting one of those jurisdictions changes the composition and weights used to form the ATT. The simulations below show over rejection when the jackknife remains defined with few treated jurisdictions, and again when few comparison jurisdictions remain.

\subsection{FastJack and TrueJack}

For DID-INT, a leave-one-jurisdiction-out replication can be implemented in two conceptually different ways. FastJack holds fixed first-stage nuisance objects that are unaffected by deleting the jurisdiction and recomputes the contrasts and aggregation steps that must change \citep{MNW-bootknife}. TrueJack is the brute-force full re-estimation jackknife: it reruns DID-INT's first stage in every leave-one-jurisdiction-out sample. TrueJack provides the reference full jackknife when deletion changes shared nuisance parameters.

The key issue is whether the first-stage CCC restrictions imply nuisance parameters that are jurisdiction-specific or shared across jurisdictions. When the first-stage parameters are jurisdiction-specific, deleting one jurisdiction need not change the nuisance components used for the remaining jurisdictions, so FastJack can be reliable. TrueJack is required when the first stage contains shared nuisance parameters that change after deleting a jurisdiction, such as pooled covariate coefficients, restrictions common across jurisdictions, restrictions common across time, or other partially pooled CCC components.

The need for full re-estimation is not monotone in flexibility. In no-covariate specifications, there may be no shared first-stage nuisance objects to re-estimate, so FastJack can be reliable. Jurisdiction-varying covariate coefficients can also permit FastJack when the omitted jurisdiction's first-stage parameters are not used for remaining jurisdictions. At the other extreme, fully flexible jurisdiction-by-time specifications can permit fast recomputation because the nuisance components for remaining jurisdiction-time cells do not depend on the omitted jurisdiction. Homogeneous covariate specifications, time-varying covariate coefficients, and other partially pooled CCC restrictions are different. In those cases, deleting a jurisdiction generally changes nuisance parameters for remaining observations, so a FastJack that holds those parameters fixed is not the jackknife for the maintained model. The first-stage coefficient structure determines which jackknife is required. With a pooled coefficient $\beta$ or time-specific coefficients $\beta_t$, deleting one jurisdiction changes coefficients used for the remaining jurisdictions, so TrueJack is required. With jurisdiction-specific coefficients $\beta_s$ or jurisdiction-by-time coefficients $\beta_{s,t}$, deletion leaves the coefficients for remaining jurisdictions unchanged, so FastJack reproduces TrueJack.

In the DID-INT simulations below, FastJack and TrueJack coincide for the remaining state-time cells because the first-stage residualization allows covariate coefficients to vary at the state-by-time level. In the UN-DID simulations, the relevant nuisance components are computed within the jurisdiction-level pieces being combined. The jackknife over-rejection documented below should therefore be read as evidence about whether deletion changes the composition and weights of the reported ATT and finite sample issues, not as evidence about the additional cost of re-estimating shared first-stage nuisance components.

\subsection{Randomization inference}

In our simulations, randomization inference (RI) treats the observed assignment as one draw from a design that fixes the numbers of early-treated, late-treated, and untreated states. It tests the sharp null of no effect for the units and treatment assignments being permuted, using the raw estimate rather than a studentized statistic \citep{MACKINNON2020435,athey2022design}. For $\ATT$, $\ATTg$, and $\ATTgt$, the reference set permutes the realized early-treated, late-treated, and untreated labels across the $H$ sampled states, holding fixed the numbers of sampled states, early-treated states, and late-treated states. For $\ATTs$, the focal state is fixed but its treatment timing may change; assignments are retained when the focal state is treated. For $\ATTst$, the focal state and calendar period are fixed; assignments are retained when the focal state is treated by that period.

For the policy-specific ATTs $\ATTp$ and $\ATTpt$, let $PG_s$ denote the joint policy-type-and-timing assignment. In the policy simulations, $PG_s$ takes values $A2$, $A4$, $B2$, $B4$, and $0$, where the letter records policy type and the number records the implementation-timing group. The RI procedure permutes these joint labels and then derives $P_s$ and $G_s$ from each reassigned $PG_s$. This keeps untreated states assigned to $P_s=G_s=0$ and rules out treated policy assignments without a treatment date. The RI $p$-value is the share of admissible reassignment estimates whose absolute value is at least as large as the observed estimate.  The simulations below evaluate the finite-sample size of this procedure. Crucially, RI tests a design-based null hypothesis that is not the same as the conventional null.

The number of admissible assignments depends on the numbers of treated and comparison states. Let \(A\) denote the total number of admissible assignments, including the observed assignment, and let \(S=A-1\) denote the number of admissible alternative assignments used to construct the randomization distribution. Let $H$ be the 
total number of jurisdictions, and $J$ be the number of early adopting jurisdictions and $L$ the number of late adopting jurisdictions, assuming two adoption cohorts. For aggregate and cohort-specific ATTs,

$$ A=\frac{H!}{J!\,L!\,(H-J-L)!}. $$
For \(\ATTs\), the number of admissible assignments is

$$ A= \frac{(H-1)!}{(J-1)!\,L!\,(H-J-L)!} + \frac{(H-1)!}{J!\,(L-1)!\,(H-J-L)!}. $$

For \(\ATTst\), the same count applies once both treatment cohorts are eligible at period \(t\); before the late treatment date, only the first term applies. When eight states are sampled and one is treated at each treatment date, there are 56 aggregate assignments, 14 state-specific assignments, and 7 early-period state-time-specific assignments. These correspond to \(S=55\), \(13\), and \(6\) alternative assignments, respectively.

When \(S\) is small, conventional randomization \(p\)-values are necessarily coarse because they depend only on the rank of the observed statistic within a small set of reassigned statistics. We therefore use the kernel-smoothing procedure of \citet{racine2007inference} when \(S<99\). The procedure estimates the distribution of the randomization statistics using a kernel-smoothed CDF, so the \(p\)-value depends on their magnitudes as well as their ranks. \citet{racine2007inference} show that this approach can improve inference when only a small number of resamples are available, and \citet{webb2023reworking} applies the same procedure to wild-cluster-bootstrap inference when the number of available bootstrap samples is small. 

In our simulations, we enumerate all admissible alternative assignments and use the kernel-smoothed \(p\)-value when \(S<99\). For \(99\le S<999\), we enumerate all alternatives and use the ordinary exact randomization \(p\)-value.  Let \(\widehat{\ATT}\) denote the observed estimate and let \(\widehat{\ATT}^{*}_s\) denote the estimate obtained under alternative assignment \(s\). Then

$$ p=\frac{1+\sum_{s=1}^{S}1(|\widehat{\ATT^*_s}| \geq |\widehat{\ATT}|)}{1+S}. $$
For \(S\ge999\), we draw 999 alternative assignments and use

$$ p=\frac{1+\sum_{s=1}^{999}1(|\widehat{\ATT^*_s}| \geq |\widehat{ATT}|)}{1000}. $$

\subsection{Software availability}

Both estimators are already available in software. UN-DID is available through the CRAN package \texttt{undidR} \citep{undidr}, with companion Stata, Julia, and Python implementations described in \citet{karim2024differenceindifferencesunpoolabledata}. DID-INT is available through the Julia package \texttt{DiDInt.jl} and the Stata wrapper \texttt{didintjl}, as described in \citet{karim2024good}. 

\section{A Worked Minimum-Wage Example}
\label{sec:minwage}

The minimum-wage application in \citet{callaway2021difference} studies county-level teen employment from 2001 through 2007, when the federal minimum wage remained fixed at \$5.15. The sample begins with states whose minimum wage equaled the federal rate; a state becomes treated when it first raises its minimum above the federal rate, producing cohorts first treated in 2004, 2006, and 2007 and a comparison group of states that never raise their minimum. These are state policy changes, not the phased federal increase that began in 2007. Because states in the same cohort set different new minimum wages, the percentage increase varies within cohorts.

We report no-covariate ATTs. Because UN-DID and DID-INT coincide in this specification, the illustration focuses on differences across levels of aggregation rather than differences between estimators. The aggregate ATT, formed by averaging across treated cohort-time cells, is $-0.052$, with a jackknife $p$-value of $0.010$ and a randomization-inference $p$-value of $0.004$.

\begin{figure}[htbp]
\centering
\begin{tikzpicture}
\begin{axis}[
width=0.78\textwidth,
height=7.2cm,
xlabel={Percent increase in statutory minimum wage},
ylabel={State-specific estimate},
xmin=0, xmax=35,
ymin=-0.12, ymax=0.03,
xtick={0,10,20,30},
ytick={-0.12,-0.08,-0.04,0,0.02},
yticklabels={$-0.12$,$-0.08$,$-0.04$,$0$,$0.02$},
grid=major,
legend style={at={(0.98,0.02)},anchor=south east,font=\footnotesize},
]
\addplot[only marks, mark=*, mark size=2.4pt, blue] coordinates {
(6.80,-0.0913341378)
};
\addlegendentry{2004}
\addplot[only marks, mark=square*, mark size=2.4pt, red] coordinates {
(10.03,-0.0040148291)
(10.48,-0.0583174415)
(12.72,-0.0740958320)
};
\addlegendentry{2006}
\addplot[only marks, mark=triangle*, mark size=2.6pt, green!60!black] coordinates {
(33.01,-0.0240445551)
(1.37,-0.0098945956)
(25.89,-0.1053778831)
(26.21,-0.0268631206)
(19.42,0.0155194789)
(19.24,0.0024166551)
(19.42,0.0000700069)
(33.01,-0.0370800547)
(12.73,-0.0070903062)
};
\addlegendentry{2007}
\addplot[densely dashed, black] coordinates {(0,0) (35,0)};
\end{axis}
\end{tikzpicture}
\caption{No-covariate state-specific ATTs plotted against each state's percentage increase in the statutory minimum wage in its implementation year. Colors and symbols denote implementation cohorts.}
\label{fig:minwage_scatter}
\end{figure}
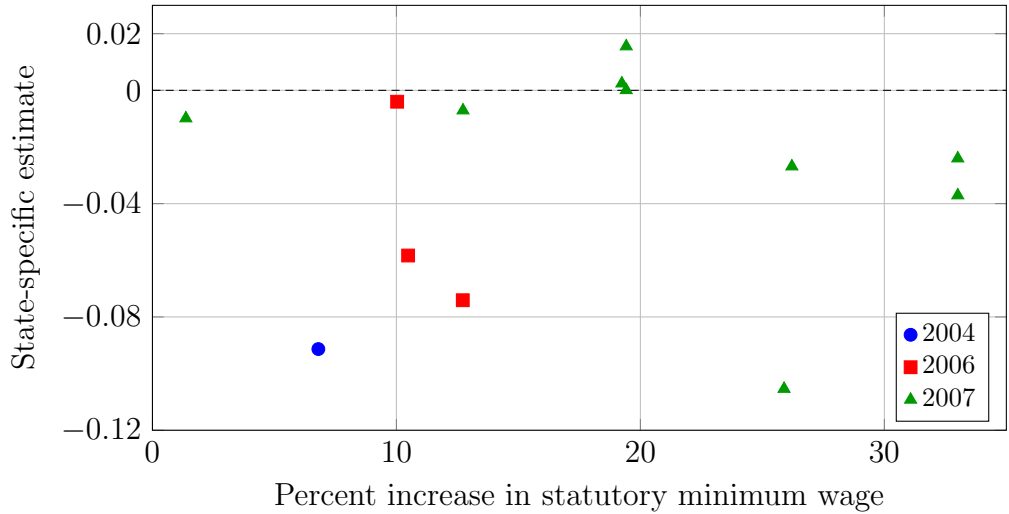

State-specific estimates reveal variation hidden by cohort aggregation.
Figure~\ref{fig:minwage_scatter} plots the no-covariate state-specific ATTs against the percentage increase in the statutory minimum wage in the implementation year. The 2007 cohort provides the clearest example: two states with minimum-wage increases of 19\% and 25\% have the most disparate estimated ATTs in the figure. The 2006 cohort shows the same pattern: states with similar statutory increases have different estimated ATTs. These within-cohort differences motivate reporting state-specific ATTs alongside cohort averages.

Within-cohort dispersion in estimated ATTs is not, by itself, evidence of treatment-effect heterogeneity: it can also arise from sampling error or violations of state-level parallel trends. A cohort-time ATT, $\ATTgt$, averages across states first treated in period $g$ when effects are measured in period $t$. When those states implement substantively different policy changes, it does not reveal which state's reform has the largest estimated effect or whether larger statutory increases are associated with larger employment responses. State-specific ATTs preserve the link between each state's policy change and its estimated employment effect.

Tables~\ref{tab:minwage_agg}--\ref{tab:minwage_attst} report the aggregate, cohort, state, and state-by-time no-covariate ATTs. UN-DID and DID-INT coincide in the no-covariate specification, so the tables report one set of estimates. Although the aggregate ATT is $-0.052$, state-level ATTs range from $-0.105$ in Michigan to $0.016$ in Montana. Only Illinois and Wisconsin have state-level RI $p$-values below $0.05$, illustrating how aggregation can obscure both the dispersion and uncertainty of state-specific estimates. For $\ATTs$ and $\ATTst$, the tables report RI $p$-values; jackknife $p$-values are unavailable.

\begin{table}[H]
\centering
\small
\begin{tabular}{lrrr}
\toprule
Estimand & Estimate & JK $p$ & RI $p$ \\
\midrule
ATT & -0.052 & 0.010 & 0.004 \\
ATT$_g$, 2004 & -0.091 & --- & 0.115 \\
ATT$_g$, 2006 & -0.047 & 0.071 & 0.148 \\
ATT$_g$, 2007 & -0.028 & 0.093 & 0.046 \\
ATT$_{g,t}$, 2004, 2004 & -0.034 & --- & 0.374 \\
ATT$_{g,t}$, 2004, 2005 & -0.071 & --- & 0.198 \\
ATT$_{g,t}$, 2004, 2006 & -0.125 & --- & 0.103 \\
ATT$_{g,t}$, 2004, 2007 & -0.136 & --- & 0.144 \\
ATT$_{g,t}$, 2006, 2006 & -0.023 & 0.377 & 0.368 \\
ATT$_{g,t}$, 2006, 2007 & -0.071 & 0.018 & 0.096 \\
ATT$_{g,t}$, 2007, 2007 & -0.028 & 0.093 & 0.046 \\
\bottomrule
\end{tabular}
\caption{Aggregate and cohort-level no-covariate ATTs in the minimum-wage application}
\label{tab:minwage_agg}
\begin{minipage}{0.95\textwidth}
\footnotesize\textit{Notes:} Jackknife $p$-values are unavailable for the 2004 cohort because Illinois is its only treated state; deleting Illinois makes the corresponding ATT undefined.
\end{minipage}
\end{table}

\begin{table}[H]
\centering
\small
\begin{tabular}{llrr}
\toprule
State & Cohort & Estimate & RI $p$ \\
\midrule
Illinois & 2004 & -0.091 & 0.037 \\
Florida & 2006 & -0.004 & 0.985 \\
Minnesota & 2006 & -0.058 & 0.627 \\
Wisconsin & 2006 & -0.074 & 0.017 \\
Colorado & 2007 & -0.024 & 0.096 \\
Maryland & 2007 & -0.010 & 0.316 \\
Michigan & 2007 & -0.105 & 0.308 \\
Missouri & 2007 & -0.027 & 0.250 \\
Montana & 2007 & 0.016 & 0.776 \\
Nevada & 2007 & 0.002 & 0.993 \\
North Carolina & 2007 & 0.000 & 1.000 \\
Ohio & 2007 & -0.037 & 0.240 \\
West Virginia & 2007 & -0.007 & 0.538 \\
\bottomrule
\end{tabular}
\caption{State-level no-covariate ATTs ($\ATTs$) in the minimum-wage application}
\label{tab:minwage_atts}
\end{table}

\begin{table}[H]
\centering
\scriptsize
\begin{tabular}{llrrr}
\toprule
State & Cohort & Year & Estimate & RI $p$ \\
\midrule
Illinois & 2004 & 2004 & -0.034 & 0.943 \\
Illinois & 2004 & 2005 & -0.071 & 0.886 \\
Illinois & 2004 & 2006 & -0.125 & 0.090 \\
Illinois & 2004 & 2007 & -0.136 & 0.021 \\
Florida & 2006 & 2006 & 0.016 & 1.000 \\
Florida & 2006 & 2007 & -0.024 & 0.602 \\
Minnesota & 2006 & 2006 & -0.021 & 0.168 \\
Minnesota & 2006 & 2007 & -0.096 & 0.029 \\
Wisconsin & 2006 & 2006 & -0.062 & 0.314 \\
Wisconsin & 2006 & 2007 & -0.087 & 0.088 \\
Colorado & 2007 & 2007 & -0.024 & 0.190 \\
Maryland & 2007 & 2007 & -0.010 & 0.296 \\
Michigan & 2007 & 2007 & -0.105 & 0.308 \\
Missouri & 2007 & 2007 & -0.027 & 0.293 \\
Montana & 2007 & 2007 & 0.016 & 0.914 \\
Nevada & 2007 & 2007 & 0.002 & 0.993 \\
North Carolina & 2007 & 2007 & 0.000 & 1.000 \\
Ohio & 2007 & 2007 & -0.037 & 0.307 \\
West Virginia & 2007 & 2007 & -0.007 & 0.505 \\
\bottomrule
\end{tabular}
\caption{State-by-time no-covariate ATTs ($\ATTst$) in the minimum-wage application}
\label{tab:minwage_attst}
\end{table}

\section{Monte Carlo Evidence}
\label{sec:mc}

\subsection{Simulation Design}

These simulations examine the size of jackknife and randomization-inference tests for aggregate and sub-aggregate average treatment effects on the treated (ATTs). The aggregate ATT combines effects across treated jurisdictions and post-treatment periods, while sub-aggregate ATTs condition on treatment cohort, jurisdiction, evaluation period, policy type, or combinations of these dimensions. We use placebo laws in Current Population Survey (CPS) data so that every null hypothesis is true by construction. The data come from the Merged Outgoing Rotation Group files for 1979--1999 and contain individual earnings, state, year, age, and education. The outcome is the log of women's weekly earnings; observations with earnings below \$20 are excluded. The resulting sample contains 547{,}818 observations. Both estimators use age, age squared, and education indicators. DID-INT also allows the first-stage covariate coefficients to vary at the state-by-time level.

Each replication draws an eight-year window and $H$ states from the CPS sample. Treatment is absorbing and assigned to an early cohort of $J$ states and a late cohort of $L$ states, with $J=L$. In the main design, the early cohort implements the placebo law in the fifth year of the window and the late cohort implements it in the seventh. The exhaustive grid is
\[
H\in\{8,12,16,20,24,28,32,36,40\},
\qquad
J=L=1,\ldots,H/2-1.
\]
Thus, every design retains at least two untreated states, and the grid contains 99 distinct $(H,J)$ cells. The results reported here use 50{,}000 independent replication families per cell.

All ATT estimands are built from jurisdiction-by-time effects, $\ATTst$, using the weights associated with the relevant aggregation. The aggregate $\ATT$ combines these effects across treated jurisdictions and post-treatment periods, whereas the sub-aggregate ATTs condition on treatment cohort ($\ATTg$ and $\ATTgt$), jurisdiction ($\ATTs$ and $\ATTst$), policy type ($\ATTp$ and $\ATTpt$), and, for the time-indexed estimands, evaluation period. The policy experiment assigns each treated state one of two policy types, A or B. Aggregating across evaluation periods gives one $\ATTp$ for each policy type. Retaining the evaluation-period dimension gives four $\ATTpt$ estimands, labeled A2, A4, B2, and B4.

We compare DID-INT and UN-DID. The jackknife deletes one sampled state at a time, recomputes the corresponding ATT estimate and any second-stage aggregation, and uses the dispersion of the leave-one-state-out estimates to form a two-sided test; deletion changes the sample used for inference but not the ATT being estimated. The jackknife is unavailable when deleting a state makes that ATT undefined. This occurs for cohort- and policy-specific ATTs when only one treated state contributes to the relevant cohort or policy type, and for state-specific ATTs whenever the focal state is deleted.

Randomization inference is implemented as described in Section~\ref{sec:estimation}, using the admissible reassignments appropriate to each ATT and the Racine--MacKinnon kernel-smoothed procedure when the number of alternative assignments is small. All rejection frequencies refer to two-sided tests at the 5 percent level.

Each replication randomly samples states and assigns treatment roles at random, so the RI results average over assignments in which smaller and larger states are treated. They do not cover settings in which treated jurisdictions are systematically smaller or larger than comparison jurisdictions. In a TWFE setting, \citet{MACKINNON2020435} show that coefficient-based RI can over-reject when treated clusters are relatively small and under-reject when they are relatively large.

\subsection{Simulation Results}

\subsubsection{Aggregate ATT}

\begin{figure}
\centering
\includegraphics[width=\textwidth,height=0.78\textheight,keepaspectratio]{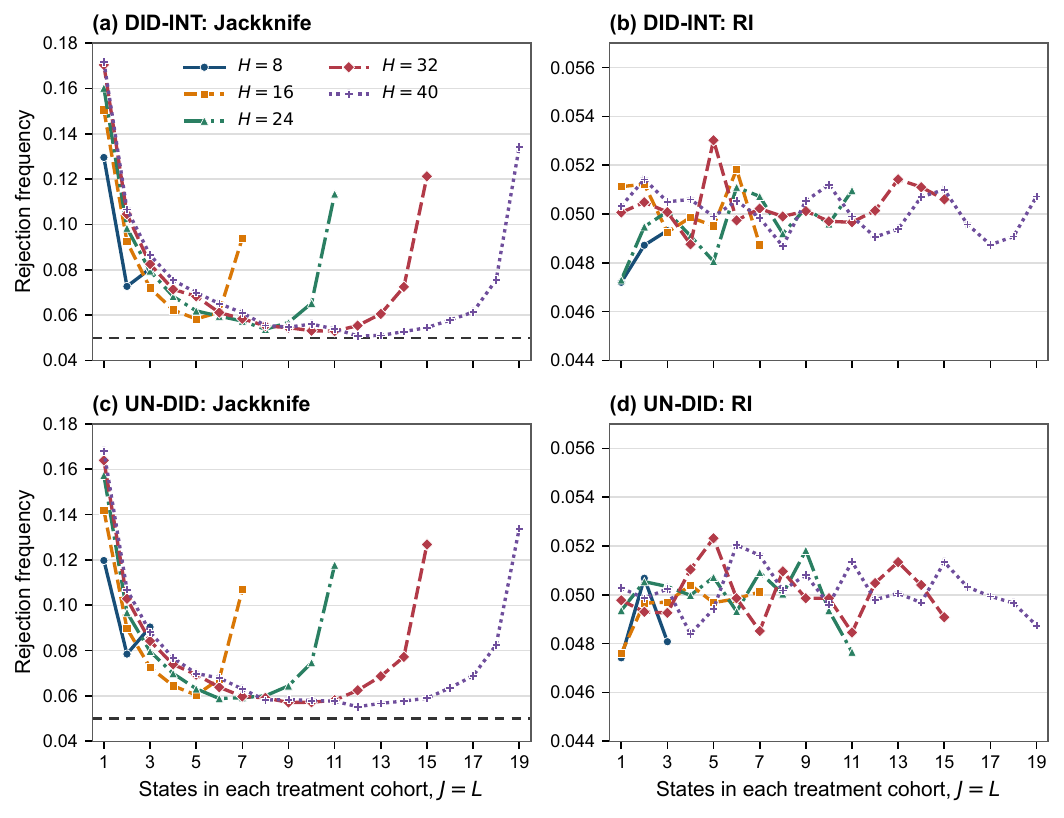}
\caption{\textbf{Aggregate ATT rejection frequencies.} The horizontal axis is $J=L$; lines select $H\in\{8,16,24,32,40\}$. Rows distinguish DID-INT and UN-DID, while columns distinguish jackknife and RI. The dashed line is the nominal 0.05 rejection rate. Jackknife panels share one vertical scale and RI panels share a separate, tighter scale. Each design cell contains 50{,}000 replication families.}
\label{fig:aggregate-att}
\end{figure}

Figure~\ref{fig:aggregate-att} shows aggregate null rejection frequencies across $J=L$ for five values of $H$. Jackknife rejection follows a U-shaped pattern: it is highest when each cohort contains only one treated state, falls as the number of treated states per cohort increases over an intermediate range, and rises again when only two comparison states remain.

At $H=40$, jackknife rejection falls from 17.18 percent to 5.60 percent between $J=1$ and $J=10$ for DID-INT, and from 16.80 percent to 5.80 percent for UN-DID. At $J=19$, when only two controls remain, rejection rises to 13.43 and 13.39 percent, respectively. The corresponding MCSEs are approximately 0.10--0.17 percentage points, so Monte Carlo sampling error cannot account for the U-shaped pattern.

Under the placebo null, RI rejection frequencies remain much closer to the 5 percent nominal level. Across the displayed cells, rejection ranges from 4.72 to 5.30 percent; across the full 99-cell grid it ranges from 4.42 to 5.30 percent. The RI panels use a tighter vertical scale because their movements are small. The MCSE near a rejection probability of 0.05 is about 0.10 percentage points, so the remaining neighboring-cell fluctuations should not be read as a systematic ranking. DID-INT and UN-DID follow the same broad pattern.

\subsubsection{\texorpdfstring{Cohort-Specific ATTs: $\ATTg$ and $\ATTgt$}{Cohort-Specific ATTs: ATT(g) and ATT(g,t)}}

\begin{figure}
\centering
\includegraphics[width=\textwidth,height=0.79\textheight,keepaspectratio]{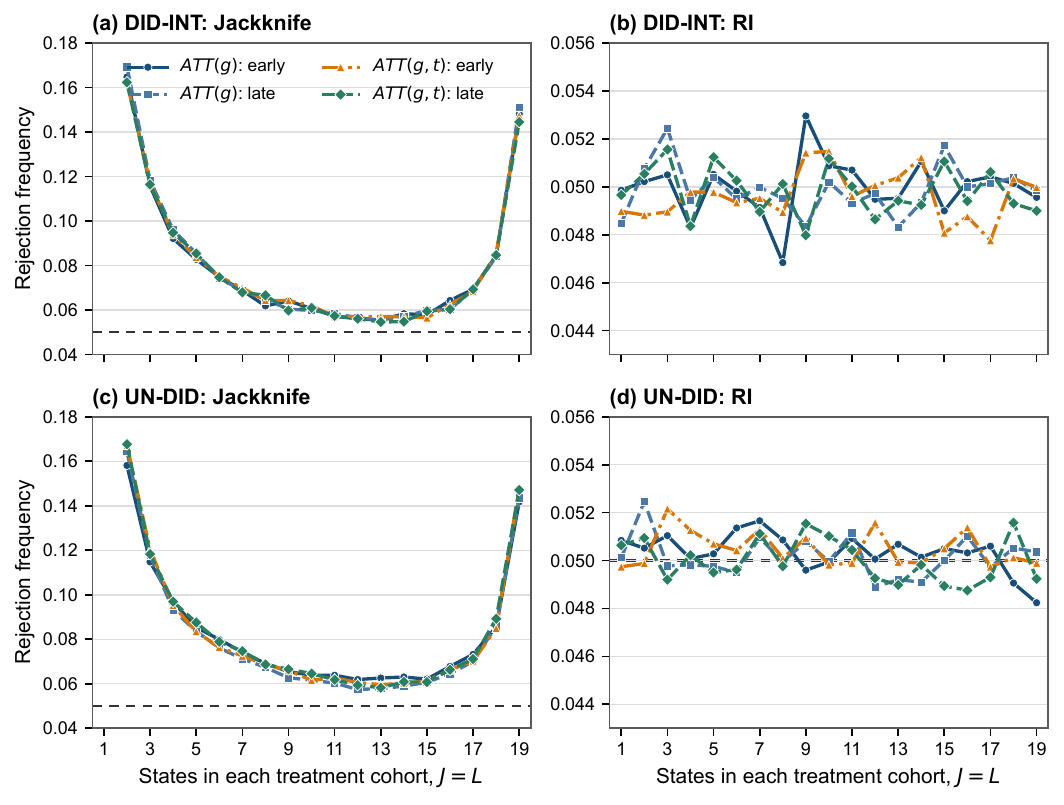}
\caption{\textbf{Cohort-specific ATT rejection frequencies.} The figure fixes $H=40$ and plots $\ATTg$ and $\ATTgt$ against $J=L$, reporting the early and late cohorts separately. Rows distinguish DID-INT and UN-DID; columns distinguish jackknife and RI. The dashed line is the nominal 0.05 rejection rate. Jackknife panels share one vertical scale and RI panels share a separate, tighter scale. When $J=1$, each cohort contains one treated jurisdiction. Deleting that jurisdiction makes its cohort-specific ATT undefined, so the leave-one-jurisdiction-out jackknife cannot be computed. Each design cell contains 50{,}000 replication families.}
\label{fig:cohort-targets}
\end{figure}

Cohort-specific ATTs are sub-aggregate ATTs that average only over treated jurisdictions in one implementation cohort. Figure~\ref{fig:cohort-targets} fixes $H=40$ and reports results separately for the early and late cohorts. The four panels compare $\ATTg$ and $\ATTgt$ under jackknife and RI for both estimators.

When each cohort contains two treated jurisdictions ($J=2$), jackknife rejection across the eight estimator, cohort, and cohort-time curves ranges from 15.81 to 16.94 percent. It falls to 5.98--6.46 percent at $J=10$, then rises to 14.18--15.11 percent when only two controls remain at $J=19$. The rejection-frequency curves for $\ATTg$ and $\ATTgt$ are often close, and neither set of tests exhibits uniformly greater size distortion. Rejection frequencies for the early and late cohorts differ by as much as 0.62 percentage points, so the figure reports them separately rather than averaging them.

Under the placebo null, RI rejection frequencies range from 4.68 to 5.30 percent across the cohort-specific panels for both DID-INT and UN-DID.

\subsubsection{\texorpdfstring{State-Specific ATTs: $\ATTs$ and $\ATTst$}{State-Specific ATTs: ATT(s) and ATT(s,t)}}

\begin{figure}[H]
\centering
\includegraphics[width=\textwidth,height=0.74\textheight,keepaspectratio]{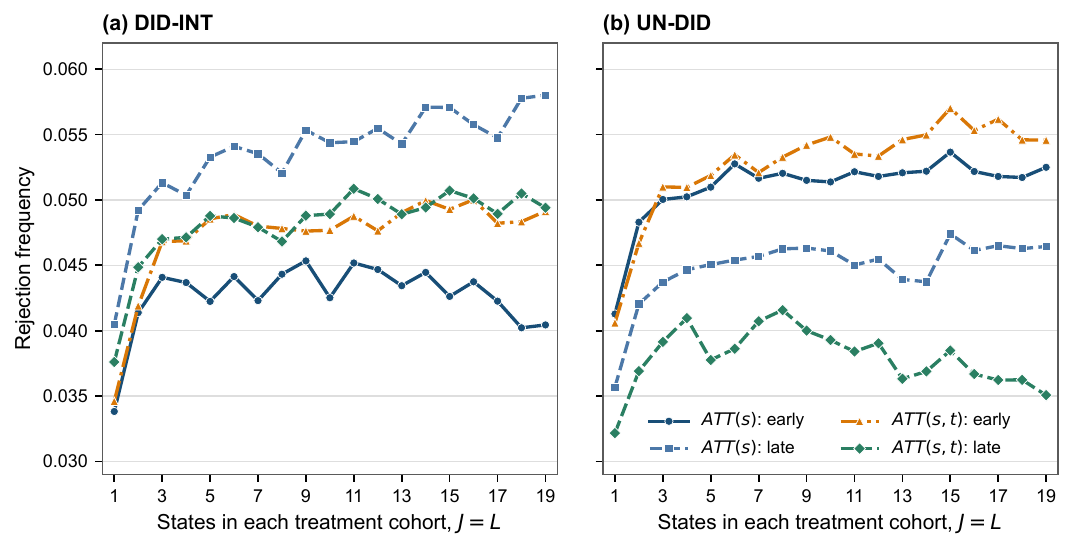}
\caption{\textbf{RI for state-specific ATTs.} The figure fixes $H=40$ and plots $\ATTs$ and $\ATTst$ against $J=L$, reporting early- and late-treated states separately. DID-INT is at left and UN-DID at right; both panels use the same tight RI scale. The dashed line is the nominal 0.05 rejection rate. Each design cell contains 50{,}000 replication families.}
\label{fig:state-targets}
\end{figure}

State-specific ATTs do not admit the usual leave-one-state-out jackknife because deleting the focal state makes the corresponding leave-one-out estimate undefined. Figure~\ref{fig:state-targets} therefore reports RI only. It fixes $H=40$ and reports results separately for early- and late-treated states and for $\ATTs$ and $\ATTst$.

Rejection ranges from 3.22 to 5.80 percent across the state panels. At $J=1$, when only one state is treated at each treatment date, all eight state- and state-time-specific rejection frequencies lie between 3.22 and 4.13 percent, indicating conservative RI tests in these cells. Rejection frequencies for early- and late-treated states also differ in some cells by substantially more than their Monte Carlo standard errors. The largest gap occurs for UN-DID $\ATTst$ at $J=17$, where the rejection frequencies for the early- and late-treated states are 5.62 and 3.62 percent. We do not have an explanation for this difference based on the assignment mechanism, so we report the early- and late-treated series separately rather than averaging them.

\subsubsection{\texorpdfstring{Policy-Specific ATTs: $\ATTp$ and $\ATTpt$}{Policy-Specific ATTs: ATT(p) and ATT(p,t)}}

\begin{figure}
\centering
\includegraphics[width=\textwidth,height=0.78\textheight,keepaspectratio]{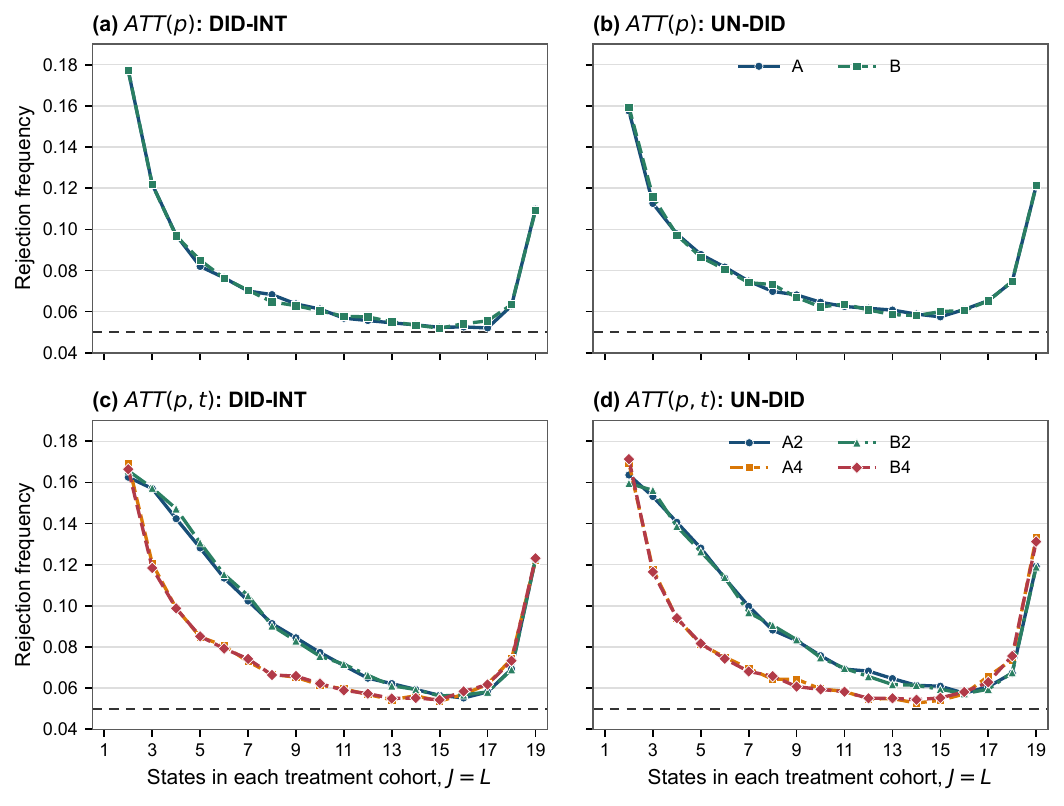}
\caption{\textbf{Jackknife rejection for policy-specific ATTs.} The figure fixes $H=40$ and varies $J=L$. The top row reports $\ATTp$ for A and B; the bottom row reports $\ATTpt$ for A2, A4, B2, and B4. DID-INT is at left and UN-DID at right. All panels share a common scale, and the dashed line is the nominal 0.05 rejection rate. When $J=1$, deleting a treated state leaves no state assigned to the relevant policy type, making the corresponding policy-specific ATT---and hence its jackknife statistic---undefined. Each design cell contains 50{,}000 replication families.}
\label{fig:policy-jk}
\end{figure}

\begin{figure}
\centering
\includegraphics[width=\textwidth,height=0.78\textheight,keepaspectratio]{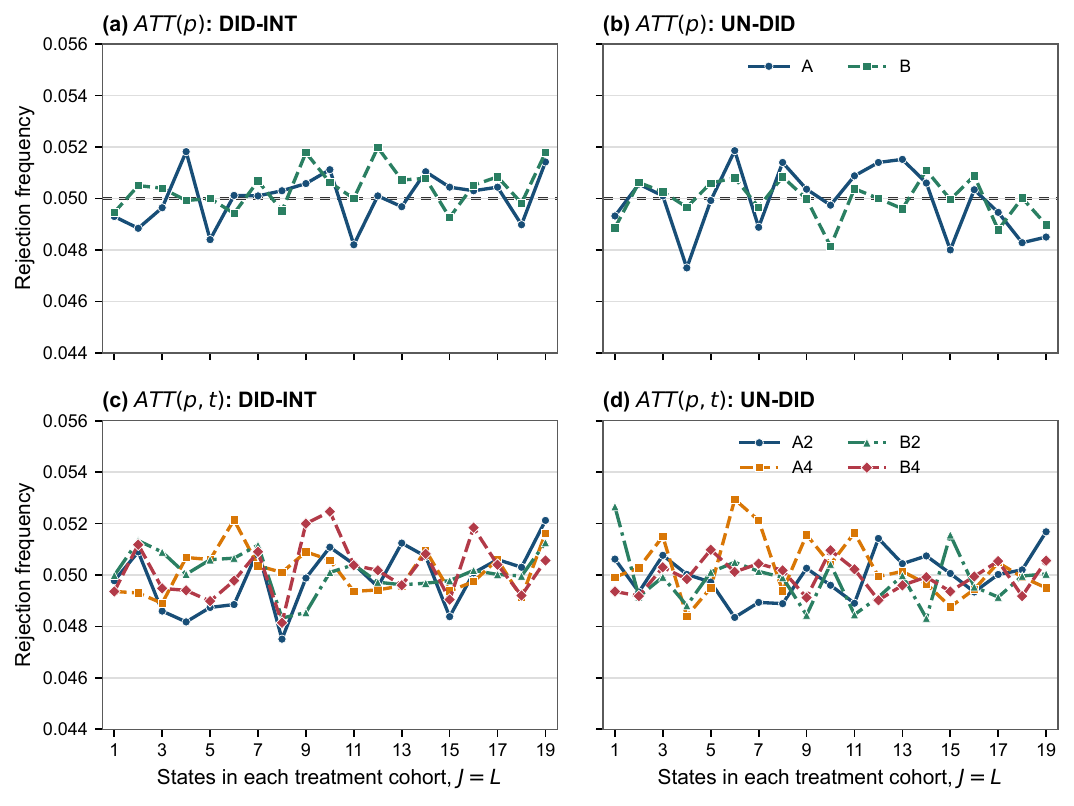}
\caption{\textbf{RI for policy-specific ATTs.} The figure fixes $H=40$ and varies $J=L$. The top row reports $\ATTp$ for A and B; the bottom row reports $\ATTpt$ for A2, A4, B2, and B4. DID-INT is at left and UN-DID at right. All panels share a tight RI scale, and the dashed line is the nominal 0.05 rejection rate. Each design cell contains 50{,}000 replication families.}
\label{fig:policy-ri}
\end{figure}

The policy experiment distinguishes the ATT for a policy type, $\ATTp$, from the ATT for a policy type at a particular evaluation period, $\ATTpt$. The top rows of Figures~\ref{fig:policy-jk} and~\ref{fig:policy-ri} report rejection frequencies for the two time-aggregated $\ATTp$ estimands, A and B, while the bottom rows report rejection frequencies for four $\ATTpt$ estimands: A2, A4, B2, and B4. Thus, A2 and A4---or B2 and B4---are distinct $\ATTpt$ estimands, not distinct $\ATTp$ estimands.

Policy-specific jackknife rejection is elevated both when only a few treated states contribute to each policy-specific ATT and when only two comparison states remain. At $J=2$, rejection across policy types and estimators ranges from 15.75 to 17.73 percent. The four rates fall to 6.04--6.46 percent at $J=10$ and rise to 10.93--12.12 percent with two controls at $J=19$. There is no stable A-versus-B ordering.

The $\ATTpt$ rejection-frequency curves share this broad U-shaped pattern, but rejection differs by evaluation period in some cells. At $J=2$, A2, A4, B2, and B4 reject between 15.95 and 17.13 percent. At $J=10$, A2 and B2 reject between 7.48 and 7.73 percent, whereas A4 and B4 reject between 5.94 and 6.22 percent. With two controls, the eight curves range from 11.90 to 13.32 percent.

Figure~\ref{fig:policy-ri} reports the corresponding RI rejection frequencies, which range from 4.73 to 5.29 percent. Differences among neighboring A/B and $\ATTpt$ curves do not reveal a stable ranking by policy type or evaluation period. In these placebo-null simulations, RI rejection frequencies vary less across policy types and evaluation periods than the jackknife rejection frequencies.

\subsubsection{Jackknife Rejection Across Levels of Aggregation}
%\enlargethispage{3\baselineskip}

\begin{figure}
\centering
\includegraphics[width=\textwidth,height=0.76\textheight,keepaspectratio]{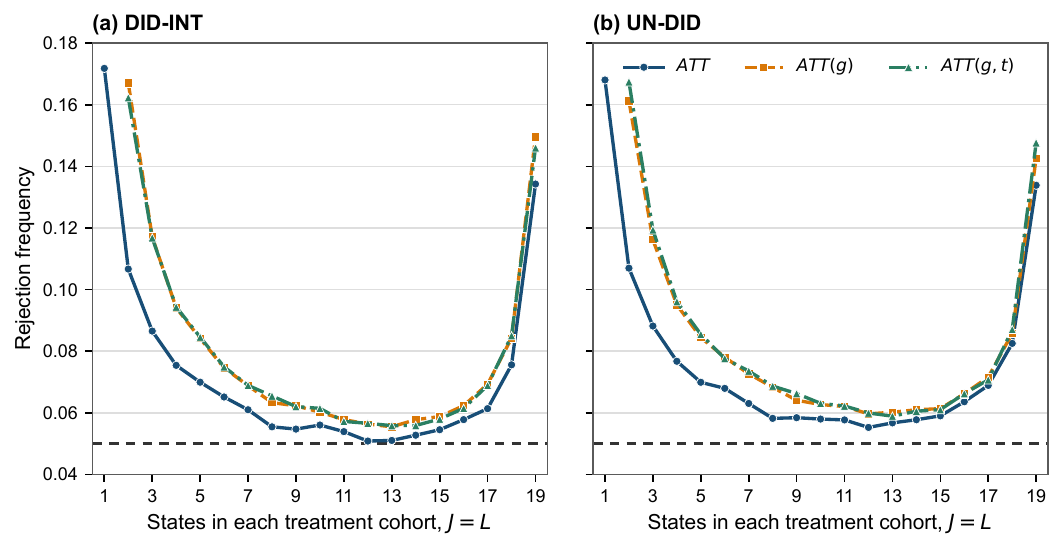}
\caption{\textbf{Jackknife rejection across levels of aggregation.} The figure fixes $H=40$ and plots rejection frequencies against $J=L$ for $\ATT$, $\ATTg$, and $\ATTgt$. DID-INT is at left and UN-DID at right; both panels use the same scale. The dashed line is the nominal 0.05 rejection rate. Rejection frequencies for $\ATTg$ and $\ATTgt$ average the corresponding early- and late-cohort frequencies. Each design cell contains 50{,}000 replication families.}
\label{fig:target-refinement}
\end{figure}

Figure~\ref{fig:target-refinement} fixes $H=40$ and compares jackknife rejection frequencies for the aggregate $\ATT$ with those for the cohort-specific $\ATTg$ and cohort-and-evaluation-period-specific $\ATTgt$. The aggregate ATT averages over both treated cohorts, whereas $\ATTg$ and $\ATTgt$ are computed separately by cohort; $\ATTgt$ additionally conditions on evaluation period. The sub-aggregate ATTs often have higher rejection frequencies when few treated states contribute to each cohort, although the three sets of tests are not strictly ordered in every cell.

At $J=2$, DID-INT rejects 10.67 percent for $\ATT$, 16.71 percent for $\ATTg$, and 16.23 percent for $\ATTgt$; the corresponding UN-DID frequencies are 10.70, 16.12, and 16.74 percent. By $J=10$, all six frequencies lie between 5.60 and 6.31 percent. With two controls at $J=19$, they rise to 13.39--14.96 percent. When few treated states contribute to each cohort, tests of the cohort-specific ATTs can overreject more than tests of the aggregate ATT, although rejection frequencies for $\ATTg$ and $\ATTgt$ are not monotonically ordered. The cohort-specific curves begin at $J=2$ because, at $J=1$, deleting a cohort's only treated state makes the corresponding leave-one-out estimate undefined.

Across these figures, jackknife size deteriorates at two distinct boundaries. Rejection is elevated when only a few treated jurisdictions contribute to the ATT being tested. It rises again when only a few comparison jurisdictions remain. Thus, jackknife performance depends on both the number of treated jurisdictions included in a given aggregate or sub-aggregate ATT and the number of comparison jurisdictions used to estimate it, not simply on the total number of sampled jurisdictions or the overall treated share.

In these placebo-null simulations, RI rejection frequencies remain much closer to 5 percent for the aggregate, cohort-specific, and policy-specific ATTs. The principal departures occur for state-specific ATTs: RI is conservative when only one state is treated at each treatment date, and some rejection frequencies for early- and late-treated states differ in other cells. This comparison concerns size under the placebo null; it does not establish a general ranking in power or computational cost. Cluster-robust inference performs most reliably with many clusters of similar size and when neither the treated nor comparison group is small \citep{MNW-guide}. These simulations include cases with few, unequal-sized clusters and with few treated or comparison jurisdictions, consistent with jackknife rejection exceeding 5 percent in many cells.  These results add to a growing literature, including \citet{mizushima2025inference} and \citet{karim2026improvedinferencecsdidusing}, showing that inferential problems long studied in TWFE settings can also arise with modern staggered-adoption DiD estimators.

\section{Conclusion}
\label{sec:concl}

Modern staggered-adoption DiD methods improve on conventional two-way fixed-effects regressions by avoiding forbidden comparisons and making the underlying treatment-effect aggregation explicit. When policy context or treatment content varies within a timing cohort, the cohort ATT averages across substantively different interventions or environments. In these settings, jurisdiction-specific ATTs should accompany the cohort average.

A cohort ATT, $\ATTg$, remains useful when jurisdictions first treated in period $g$ implement comparable policies in comparable environments. Otherwise, it should be interpreted as a weighted average across different interventions rather than the effect of a common policy. Jurisdiction-specific estimates within each cohort can reveal dispersion that cohort averages conceal. That dispersion may reflect differences in policy design or economic environment, sampling error, or treatment-effect heterogeneity.
It may also reflect bias from violations of jurisdiction-level parallel trends. Choosing whether to aggregate by jurisdiction, policy type, or cohort therefore requires substantive knowledge of policy design and the environments in which policies are implemented. Applied work should report jurisdiction-specific ATTs when policy design varies within cohorts.

Jurisdiction-specific ATTs, $\ATTst$ and $\ATTs$, link each reported effect to a particular jurisdiction, policy element, or context. Their identifying assumptions---including parallel trends, no anticipation, and no cross-jurisdiction spillovers---must hold at the jurisdiction rather than cohort level.
Both UN-DID and DID-INT estimate jurisdiction-by-time ATTs that can be aggregated to the cohort, jurisdiction, policy-type, and overall levels. In the no-covariate minimum-wage application, the two estimators yield the same estimates.

The simulations yield several practical conclusions about inference for aggregate and sub-aggregate ATTs. RI rejection frequencies remain close to 5 percent for most aggregate, cohort-, state-, and policy-specific ATTs. Some state-specific tests are conservative when only one jurisdiction implements the placebo policy at each treatment date. RI tests a design-based null rather than the null underlying conventional large-sample inference. When the jackknife remains defined, it follows a two-boundary pattern: it over-rejects when only a few treated jurisdictions contribute to the ATT and again when only a few comparison jurisdictions remain. The usual leave-one-jurisdiction-out jackknife is unavailable for jurisdiction-specific ATTs. When a jackknife is available, FastJack reproduces full re-estimation only if deleting a jurisdiction leaves the first-stage estimates for the remaining jurisdictions unchanged.

Methodological work should develop diagnostics for jurisdiction-level parallel trends and design-based RI procedures with better finite-sample performance when only a few jurisdictions implement the policy. The power of RI tests for jurisdiction-specific ATTs remains an important open question, especially when only a few jurisdictions implement the policy. Progress on jurisdiction-level diagnostics, finite-sample RI, and power comparisons would make staggered-adoption DiD more useful for comparing policies that differ in design or implementation context.

\clearpage

\noindent \textbf{Statement on Use of AI}

\noindent Generative AI was used extensively in preparing this manuscript.  The ideas for the estimators, the inference procedures, the monte carlo design, and the empirical examples all came from the authors.  The literature review is also entirely done by the authors.  ChatGPT helped organize dictated remarks and past beamer slides into a skeletal rough draft.  Codex combined disparate code fragments into a unifed monte carlo package, which was then optimized by incorporating code provided by the authors, and suggestions from the authors and Codex.  An ensemble of Refine, Codex, Grok, and ChatGPT helped with editing the prose, and still left plenty of errors and ambiguities for the authors to correct and improve themselves with human labour.  These tools were mostly used because they are enjoyable to work with,  write very elegant code, and to simultaneously improve productivity and reproducibility. 

\noindent \textbf{Data Availability Statement}

The CPS Merged Outgoing Rotation Group earnings files used in the simulations are publicly available from the National Bureau of Economic Research at \url{https://www.nber.org/research/data/current-population-survey-cps-merged-outgoing-rotation-group-earnings-data}. The county-level teen-employment data used in the minimum-wage application are the replication data for Callaway and Sant'Anna (2021), available at \url{https://doi.org/10.1016/j.jeconom.2020.12.001}; a public replication copy is hosted at \url{https://github.com/pedrohcgs/CS_RR}. State statutory minimum-wage levels used to calculate implementation-year percentage changes come from David Neumark's State Minimum Wage Data Set through September 2019, available at \url{https://sites.socsci.uci.edu/~dneumark/datasets.html}. 

\clearpage

\bibliographystyle{cjemod}
\bibliography{mybib}

@article{athey2022design,
  author  = {Athey, Susan and Imbens, Guido W.},
  title   = {Design-Based Analysis in Difference-in-Differences Settings with Staggered Adoption},
  journal = {Journal of Econometrics},
  year    = {2022},
  volume  = {226},
  number  = {1},
  pages   = {62--79}
}

@article{AustinApold2023,
  author  = {Austin, Nichole and Apold, Victoria},
  title   = {Canadian Access to Assisted Reproduction: Mapping Changes Over Time},
  journal = {Journal of Obstetrics and Gynaecology Canada},
  year    = {2023},
  volume  = {45},
  number  = {9},
  pages   = {644--645}
}

@article{bertrand2004much,
  author  = {Bertrand, Marianne and Duflo, Esther and Mullainathan, Sendhil},
  title   = {How Much Should We Trust Differences-in-Differences Estimates?},
  journal = {The Quarterly Journal of Economics},
  year    = {2004},
  volume  = {119},
  number  = {1},
  pages   = {249--275}
}

@article{callaway2021difference,
  author  = {Callaway, Brantly and Sant'Anna, Pedro H. C.},
  title   = {Difference-in-Differences with Multiple Time Periods},
  journal = {Journal of Econometrics},
  year    = {2021},
  volume  = {225},
  number  = {2},
  pages   = {200--230}
}

@article{conley_2011,
  author  = {Conley, Timothy G. and Taber, Christopher R.},
  title   = {Inference with ``Difference in Differences'' with a Small Number of Policy Changes},
  journal = {The Review of Economics and Statistics},
  year    = {2011},
  volume  = {93},
  number  = {1},
  pages   = {113--125}
}

@article{de2020twott,
  author  = {De Chaisemartin, Cl{\'e}ment and D'Haultfoeuille, Xavier},
  title   = {Two-Way Fixed Effects Estimators with Heterogeneous Treatment Effects},
  journal = {American Economic Review},
  year    = {2020},
  volume  = {110},
  number  = {9},
  pages   = {2964--2996}
}

@article{goodman2021difference,
  author  = {Goodman-Bacon, Andrew},
  title   = {Difference-in-Differences with Variation in Treatment Timing},
  journal = {Journal of Econometrics},
  year    = {2021},
  volume  = {225},
  number  = {2},
  pages   = {254--277}
}

@Manual{undidr,
  author = {Jamieson, Eric},
  title  = {undidR: Difference-in-Differences with Unpoolable Data},
  year   = {2025},
  note   = {R package version 1.0.2}
}

@article{karim2024good,
  author  = {Karim, Sunny and Webb, Matthew D.},
  title   = {Good Controls Gone Bad: Difference-in-Differences with Covariates},
  journal = {arXiv:2412.14447},
  year    = {2024}
}

@article{karim2024differenceindifferencesunpoolabledata,
  author  = {Karim, Sunny and Webb, Matthew D. and Austin, Nichole and Strumpf, Erin},
  title   = {Difference-in-Differences with Unpoolable Data},
  journal = {arXiv:2403.15910},
  year    = {2024}
}

@article{karim2026improvedinferencecsdidusing,
  author  = {Karim, Sunny and Nielsen, Morten {\O}rregaard and MacKinnon, James G. and Webb, Matthew D.},
  title   = {Improved Inference for {CSDID} Using the Cluster Jackknife},
  journal = {arXiv:2602.12043},
  year    = {2026}
}

@article{MNW-bootknife,
  author  = {MacKinnon, James G. and Nielsen, Morten {\O}. and Webb, Matthew D.},
  title   = {Fast and Reliable Jackknife and Bootstrap Methods for Cluster-Robust Inference},
  journal = {Journal of Applied Econometrics},
  year    = {2023},
  volume  = {38},
  pages   = {671--694}
}

@article{MNW-guide,
  author  = {MacKinnon, James G. and Nielsen, Morten {\O}. and Webb, Matthew D.},
  title   = {Cluster-Robust Inference: {A} Guide to Empirical Practice},
  journal = {Journal of Econometrics},
  year    = {2023},
  volume  = {232},
  pages   = {272--299}
}

@article{mackinnon2017wild,
  author  = {MacKinnon, James G. and Webb, Matthew D.},
  title   = {Wild Bootstrap Inference for Wildly Different Cluster Sizes},
  journal = {Journal of Applied Econometrics},
  year    = {2017},
  volume  = {32},
  number  = {2},
  pages   = {233--254}
}

@article{MACKINNON2020435,
  author  = {MacKinnon, James G. and Webb, Matthew D.},
  title   = {Randomization Inference for Difference-in-Differences with Few Treated Clusters},
  journal = {Journal of Econometrics},
  year    = {2020},
  volume  = {218},
  number  = {2},
  pages   = {435--450}
}

@article{mikola2023finish,
  author  = {Mikola, Derek and Webb, Matthew D.},
  title   = {Finish It and It Is Free: An Evaluation of College Graduation Subsidies},
  journal = {Economics of Education Review},
  year    = {2023},
  volume  = {93},
  pages   = {102355}
}

@article{sun2021estimating,
  author  = {Sun, Liyang and Abraham, Sarah},
  title   = {Estimating Dynamic Treatment Effects in Event Studies with Heterogeneous Treatment Effects},
  journal = {Journal of Econometrics},
  year    = {2021},
  volume  = {225},
  number  = {2},
  pages   = {175--199}
}

@article{borusyak2024revisiting,
  title={Revisiting event-study designs: robust and efficient estimation},
  author={Borusyak, Kirill and Jaravel, Xavier and Spiess, Jann},
  journal={Review of Economic Studies},
  volume={91},
  number={6},
  pages={3253--3285},
  year={2024},
  publisher={Oxford University Press UK}
}

@article{hansen2025standard,
  author  = {Hansen, Bruce E.},
  title   = {Standard Errors for Difference-in-Difference Regression},
  journal = {Journal of Applied Econometrics},
  year    = {2025},
  volume  = {40},
  number  = {3},
  pages   = {291--309},
  doi     = {10.1002/jae.3110}
}

@article{hansen2025jackknife,
  author  = {Hansen, Bruce E.},
  title   = {Jackknife Standard Errors for Clustered Regression},
  journal = {Review of Economic Studies},
  year    = {2025},
  note    = {Forthcoming}
}

@article{MNW-summclust,
  author  = {MacKinnon, James G. and Nielsen, Morten {\O}rregaard and Webb, Matthew D.},
  title   = {Leverage, Influence, and the Jackknife in Clustered Regression Models:
             Reliable Inference Using summclust},
  journal = {Stata Journal},
  year    = {2023},
  volume  = {23},
  number  = {4},
  pages   = {942--982},
  doi     = {10.1177/1536867X231212433}
}

@article{webb2023reworking,
  author  = {Webb, Matthew D.},
  title   = {Reworking Wild Bootstrap-Based Inference for Clustered Errors},
  journal = {Canadian Journal of Economics/Revue canadienne d'{\'e}conomique},
  year    = {2023},
  volume  = {56},
  number  = {3},
  pages   = {839--858},
  doi     = {10.1111/caje.12661}
}

@article{racine2007inference,
  author  = {Racine, Jeffrey S. and MacKinnon, James G.},
  title   = {Inference via Kernel Smoothing of Bootstrap {P} Values},
  journal = {Computational Statistics \& Data Analysis},
  year    = {2007},
  volume  = {51},
  number  = {12},
  pages   = {5949--5957},
  doi     = {10.1016/j.csda.2006.11.013}
}

@techreport{mizushima2025inference,
  title={Inference with modern difference-in-differences methods},
  author={Mizushima, Yuji and Powell, David},
  institution={SSRN 5221387},
  year={2025}
}

\end{document}